\documentclass[prd]{revtex4}
\usepackage{float}
\usepackage{graphicx}
\usepackage{caption}
\usepackage{amsmath}
\usepackage{array}
\usepackage{bm}
\usepackage{amssymb}
\usepackage{amsfonts}
\usepackage{amssymb}
\usepackage{color}

\begin{document}
\title{A Comparative Study of Mass Extraction Schemes and $\pi^\pm-\rho^\pm$ Mixing}
\author{Ziyue Wang}
\affiliation{School of Physics and Optoelectronic Engineering, Beijing University of Technology, Beijing 100021, China}
\date{\today}
\begin{abstract}
We study the origin of the non-monotonic magnetic-field dependence of the lowest charged pion excitation observed in lattice QCD. In a magnetic field, the charged pion mixes with the longitudinally polarized charged rho meson, which shares the same quantum numbers. Within the SU(2)$_f$ Nambu–Jona-Lasinio model supplemented by a gauge invariant tree-level $\pi-\rho$ mixing operator constrained by the experimental $\rho^\pm\rightarrow\pi^\pm\gamma$ decay width, we compare four mass-extraction schemes: rest-mass reconstruction, local bosonization, direct determinant solving with Landau projection, and near-pole expansion. The rest-mass scheme cannot reproduce the lattice-type turnover, while in the local derivative-expansion scheme the turnover presence but is weak which occurs at large magnetic field. By contrast, the direct determinant and near-pole schemes both retain a robust non-monotonic lowest mode. The former is most faithful to the Landau-level kinematics of the charged excitation, while the latter most clearly shows that residue suppression enhances the effective mixing after canonical normalization. Our results indicate that the lattice behavior is a genuine quasiparticle mixing effect, but one whose robustness depends crucially on how the charged-meson pole structure is extracted in a magnetic field.
\end{abstract}

\maketitle

\section{INTRODUCTION}
External magnetic fields provide a useful probe of hadronic structure because they couple directly to electric charge and spin, and therefore resolve aspects of QCD bound states that are hidden in vacuum. Such fields are relevant in several physical settings, ranging from the early universe and magnetars to the short-lived but extremely intense fields generated in non-central heavy-ion collisions~\cite{Grasso:2000wj,Durrer:2013pga,Kiuchi:2015sga,Skokov:2009qp,Deng:2012pc}. When the magnetic scale becomes comparable with typical hadronic scales, the response of QCD matter is no longer perturbative in a simple electromagnetic sense. The magnetic field can affect chiral dynamics, modify quark and hadron dispersion relations, and change the internal structure of mesonic bound states~\cite{Kharzeev:2007jp,Kharzeev:2010gr,Gursoy:2014aka,Gusynin:1995nb,Kharzeev:2010gd,Bali:2012zg,Tomiya:2019nym,Bali:2014kia,Bali:2012jv}. Reviews of QCD in magnetic backgrounds can be found in Refs.~\cite{Miransky:2015ava,Fukushima:2012kc,Andersen:2014xxa}.

One of the most direct manifestations of the magnetic field is its effect on charged hadrons. For a pointlike charged particle, the transverse motion is quantized into Landau levels, and the lowest charged excitation is expected to grow roughly with the magnetic scale. Hadrons, however, are composite objects. Their correlation functions also contain information about binding, polarization, wave-function normalization, and possible mixing with other modes. Therefore the charged meson spectrum in a magnetic field can not be described by a simple lowest Landau level dispersion relation.

Lattice QCD calculations have provided important nonperturbative information on this problem~\cite{Bali:2011qj,Bali:2012zg,Ilgenfritz:2013ara,Ding:2020hxw,Ding:2020jui,Ding:2026qzu}. A particularly striking result is that the lowest charged-pion excitation shows a non-monotonic dependence on the magnetic field. Instead of increasing monotonically as suggested by the naive lowest-Landau-level expectation, the energy rises at weak field, reaches a maximum at intermediate field strength, and then decreases as the magnetic field becomes stronger. Similar behavior has also been observed for charged kaons. This qualitative feature indicates that the lattice observable is sensitive not only to Landau quantization, but also to the internal dynamics of the charged meson in the magnetic background.

The magnetic-field dependence of light mesons has been studied in a variety of effective approaches, including NJL-type models, functional methods, and holographic models~\cite{Inagaki:2003yi,Yu:2014xoa,Fayazbakhsh:2014mca,Coppola:2018vkw,Coppola:2019uyr,Chaudhuri:2019lbw,Kamikado:2013pya,Mamo:2015dea,Li:2016gfn}. The charged vector meson sector has also been widely discussed, especially because the magnetic field separates different spin projections and can strongly affect the charged $\rho$ channel~\cite{Liu:2014uwa,Ghosh:2016evc,Kawaguchi:2015gpt}. More recent works have explored several possible sources of the lattice trend, including magnetic-field-dependent effective couplings~\cite{Avancini:2021pmi}, meson mixing~\cite{Carlomagno:2022arc,Coppola:2023mmq}, wave-function renormalization effects~\cite{Wen:2023qcz}, and inverse magnetic catalysis~\cite{Li:2023rsy}. These studies have clarified many aspects of meson spectroscopy in a magnetic field. Nevertheless, reproducing the lattice-type turnover in a controlled and robust way remains nontrivial.

A natural mechanism is the mixing between the charged pion and the longitudinally polarized charged rho meson. In vacuum, the pseudoscalar pion and vector rho meson are distinct excitations. In a magnetic field, however, rotational symmetry is reduced and the $s_z=0$ component of the charged rho meson can have the same residual quantum numbers as the charged pion. The magnetic background can therefore induce $\pi^\pm-\rho^\pm$ mixing. Such mixing generates level repulsion between pion-like and rho-like modes and may lower the lightest eigenmode at sufficiently large magnetic field. In our previous work~\cite{Wang:2026xsm}, we focused on this dynamical mechanism and showed how $\pi-\rho$ mixing can provide a possible origin of the non-monotonic charged meson energy observed in lattice QCD.

The present work addresses a different but closely related issue. Once a magnetic field is present, the word “mass” is no longer unique. In vacuum, Lorentz symmetry ties together the pole mass, the rest energy, and the mass parameter appearing in a low-energy effective Lagrangian. In a magnetic background this equivalence is lost. Charged excitations are described by Landau-level eigenstates rather than ordinary plane waves, longitudinal and transverse dynamics are separated, and the residues of composite meson poles may depend strongly on the magnetic field. As a result, different operational procedures for extracting a meson mass can lead to different magnetic-field dependences, even when they start from the same microscopic dynamics.

For this reason, the main goal of this paper is to compare mass-extraction schemes for the coupled $\pi^\pm-\rho^\pm$ system. We work within a common SU(2)$_f$ Nambu--Jona-Lasinio framework and include a gauge-invariant $\pi-\rho$ mixing operator constrained by the physical $\rho^\pm\to\pi^\pm\gamma$ decay width. On this common basis, we examine four schemes: the rest-mass construction from the RPA pole at zero spatial momentum, the local bosonized derivative expansion, the direct solution of the Landau-projected determinant, and the near-pole quasiparticle expansion with canonical normalization. Since the same microscopic kernel is used throughout, the comparison isolates the role played by Landau-level projection, locality, and wave-function normalization in the extracted charged spectrum.

We find that the four schemes form a clear hierarchy. The rest-mass construction fails to reproduce the lattice-type turnover in the physical lowest-Landau-level energy, because it first extracts a vacuum-like rest mass and only afterwards reconstructs the charged energy. The local derivative-expansion scheme can generate a turnover, but the effect is weak and appears only at relatively large magnetic field, suggesting that a strictly local expansion does not fully preserve the charged pole structure relevant for the lattice observable. By contrast, the direct determinant and near-pole schemes both retain a robust non-monotonic lowest mode. The direct determinant scheme gives the most faithful treatment of the charged Landau eigenmode, while the near-pole scheme makes transparent that suppression of mesonic residues enhances the effective mixing after canonical normalization.

Our results therefore refine the physical interpretation of the lattice behavior. The turnover is not merely a generic consequence of adding a $\pi-\rho$ mixing term to an effective Lagrangian. It appears robustly only when the charged meson is treated as a Landau-level quasiparticle and when its pole normalization is handled consistently. In this sense, the lattice result probes not only mass shifts, but also the magnetic-field dependence of the charged pole structure and the quasiparticle residues.

This paper is organized as follows. Sec.~\ref{definition} introduces the mass definitions in a magnetic field and outlines the four extraction schemes considered here. Sec.~\ref{com_input} presents the common microscopic input, including the NJL model, the $\pi^\pm-\rho^\pm_{s_z=0}$ quadratic kernel, and the vacuum matching of the effective mixing term. Sec.~\ref{sec_rest}--\ref{sec_nearpole} implement the rest-mass, local bosonized, direct determinant, and near-pole extraction schemes, respectively. Sec.~\ref{sec_compare} compares the four results, while Sec.~\ref{sec_implic} discusses their implications for lattice observables and effective descriptions. Sec.~\ref{sec_outlook} gives the outlook.

\section{Mass definitions in a magnetic field and extraction strategies}
\label{definition}
In vacuum, Lorentz invariance ensures that different operational definitions of hadron mass coincide. The pole of the propagator, the rest energy of the excitation, and the mass parameter appearing in an effective Lagrangian all refer to the same physical quantity. In a magnetic field, however, this equivalence is lost. Rotational and Lorentz symmetries are explicitly broken, charged excitations occupy discrete Landau levels, and temporal and spatial components of correlators no longer enter on the same footing. As a result, the notion of “mass” becomes scheme dependent.

In the present work, we compare four extraction strategies that arise naturally in effective QCD treatments of charged mesons in a magnetic background. All four are derived from the same microscopic $\pi-\rho$ quadratic kernel, but they differ in how they treat Landau-level kinematics, locality, and wave-function renormalization. Rather than viewing them as competing models, we use them as complementary projections of the same underlying dynamics.

\paragraph{Rest-mass extraction (RPA Pole at Zero Momentum)}
The most direct extension of the vacuum pole definition is to determine the mass from the condition
\begin{eqnarray}
\text{det}\mathcal{K}(q_0,\vec{q}=0)=0
\end{eqnarray}
where $\mathcal{K}$ is the quadratic kernel in the meson sector. This definition answers the question: \textit{What is the energy required to create a composite excitation at rest?}

In this scheme, the magnetic field enters only through the quark loop polarization functions. No explicit Landau level structure is imposed at the meson level. The resulting quantity is therefore a rest-frame pole of the composite correlator, rather than the energy of a propagating charged quasiparticle. This construction provides the simplest baseline for comparison, since it most closely parallels the familiar vacuum RPA pole definition~\cite{Klevansky:1992qe, Hatsuda:1994pi, Avancini:2022qcp}.

\paragraph{Local extraction from bosonization}
An alternative viewpoint is obtained by bosonizing the underlying quark model and performing a derivative expansion of the effective action~\cite{Klevansky:1992qe, Wang:2017vtn, Coppola:2019uyr}. In this approach, the nonlocal quadratic kernel is reorganized into a local mesonic theory with magnetic field dependent kinetic and mass coefficients,
\begin{eqnarray}
\mathcal{L}_\text{eff}=Z_\phi(B)|D_\mu\phi|^2-m_\phi^2(B)|\phi|^2+\cdots.
\end{eqnarray}
This scheme answers a different question: what effective mass parameter governs the charged meson in a local mesonic description? The magnetic field is incorporated through the covariant derivative, so Landau quantization is introduced at the level of the effective meson fields. However, the coefficients are extracted from a derivative expansion around small momentum, rather than from the physical pole itself. The local scheme therefore provides an EFT reinterpretation of the same microscopic dynamics, but one that relies on locality and small-momentum expansion.

\paragraph{Direct Determinant with Landau Projection}
For a charged excitation in a magnetic field, the physically relevant mode is not a plane wave but a Landau-level eigenstate. A more kinematically faithful definition is therefore obtained by projecting the quadratic kernel onto a fixed Landau level and solving
\begin{eqnarray}
\text{det}\mathcal{K}_{LLL}(q_0)=0,
\end{eqnarray}
This scheme answers the question: what is the energy of the lowest Landau eigenmode of the coupled $\pi-\rho$ system? Unlike the rest-mass construction, Landau quantization is imposed directly on the external meson state \cite{Coppola:2018vkw, Li:2020hlp, Carlomagno:2022arc, GomezDumm:2023owj}. Unlike the local expansion, the pole is extracted from the full projected kernel rather than from a derivative expansion. This makes the method the most direct microscopic determination of the charged excitation energy in a magnetic background.

\paragraph{Near-Pole Extraction (Quasiparticle effective theory)}
Finally, one may expand the Landau-projected quadratic kernel around the unmixed physical poles and canonically normalize the propagating modes \cite{Xu:2022kng, Cao:2023syu, Wang:2026xsm},
\begin{eqnarray}
\mathcal{K}_{LLL}(q_0)\approx Z_\phi(B)(q_0^2-E_\phi^2(B)),
\end{eqnarray}
In this form, the residues $Z_\phi(B)$ enter explicitly and the mixing term is naturally expressed in terms of canonically normalized fields.

This scheme answers the question: what is the mass of the physical quasiparticle eigenmode once pole normalization is taken into account? Since the effective mixing scales inversely with the square root of the residues, the near-pole formulation makes transparent how suppression of the mesonic residues can enhance level repulsion. Conceptually, it is not a different kinematics from the direct determinant approach, but a quasiparticle reorganization of the same Landau-projected pole structure.

The four strategies thus form a natural hierarchy. The rest-mass scheme is the simplest vacuum-inspired construction; the local bosonized scheme recasts the problem into a local effective theory; the direct determinant method solves the Landau-projected charged mode directly; and the near-pole expansion provides the most transparent quasiparticle interpretation of the same pole structure. In the following sections, we implement each strategy explicitly and compare their predictions for the coupled $\pi^\pm-\rho^\pm$ system in a magnetic field. Before doing so, we first summarize the common microscopic input shared by all four schemes.

\section{Common Microscopic Input}
\label{com_input}
\paragraph{NJL model in a magnetic field}
The Lagrangian of the SU(2) Nambu-Jona-Lasinio (NJL) model is given by 
\begin{eqnarray}
\mathcal{L}=\bar{\psi}(iD\!\!\!\!/-\hat{m})\psi+G_S[(\bar{\psi}\psi)^2+(\bar{\psi}i\gamma^5\psi)^2]
-G_V[(\bar{\psi}\gamma^\mu\tau^a\psi)^2],
\end{eqnarray}
where $\psi$ corresponds to the quark field of two light flavor u and d, $\hat{m}$=diag$(m_u,m_d)$ is the current quark mass matrix of u and d quarks, $\tau^a=(I,\vec{\tau})$ with $\vec{\tau}=(\tau^1,\tau^2,\tau^3)$ representing the isospin Pauli matrices, and $G_S$ and $G_V$ are the coupling constants with respect to the scalar (pseudoscalar) and the vector (axial-vector) channels \cite{Klevansky:1992qe, Hatsuda:1994pi, Vogl:1991qt}. The covariant derivative, $D_\mu=\partial_\mu-iq_fA_\mu^{ext}$ couples quarks to an external magnetic field $\vec{B}=(0,0,B)$ along the positive $z$ direction via a background field, with Landau gauge $A=-By\delta_1^k$. $q_f=(-1/3,2/3)$ is the electric charge of the quark field. The field strength tensor $F_{\mu\nu}$ is defined by $F_{\mu\nu}=\partial_{[\mu}A_{\nu]}^{ext}$. 

After Hubbard–Stratonovich transformation we have quarks coupled linearly to auxiliary meson fields
\begin{eqnarray}
\mathcal{L}=\bar{\psi}(x)(i\gamma^\mu D_\mu-m_0)\psi(x)
-\bar{\psi}(\sigma+i\gamma_5\vec{\tau}\cdot\vec{\pi})\psi
-\frac{(\sigma^2+\vec{\pi}^2)}{4G_S}
+\frac{(\rho_\mu^a \rho_{a\mu})}{4G_V},
\end{eqnarray}
where the Euler-Lagrange equations of motion for the auxiliary fields lead to the constraints
\begin{eqnarray}
\sigma(x) &=& -2G_S\bar{\psi}(x)\psi(x),\nonumber\\
\vec{\pi}(x) &=&  -2G_S\bar{\psi}(x)i\gamma_5\vec{\tau}\psi(x),\nonumber\\
\rho_\mu^a(x) &=& -2G_V\bar{\psi}(x)\gamma_\mu\tau^a\psi(x).
\end{eqnarray}
For the quark propagator, we take the Landau-level representation \cite{Schwinger:1951nm, Miransky:2015ava} $S_{f}(r,r')=e^{i\Phi(\mathbf{r}_\perp,\mathbf{r}'_\perp)}S_{f}(r-r')$, where $\Phi(\mathbf{r}_\perp,\mathbf{r}'_\perp)$ is the Schwinger phase defined by $\Phi(\mathbf{r}_\perp,\mathbf{r}'_\perp)=-\frac{s_\perp^f(x-x')(y+y')|q_fB|}{2}$, with $s_\perp^f=\text{sign}(q_fB)$, $q_f$ is the charge of the fermion field. The translation invariant part of the propagator can be transformed to momentum space $S_{f}(r-r')=\int\frac{d^4p}{(2\pi)^4}e^{-ip\cdot(r-r')}{S}_{f}(p)$, with 
\begin{eqnarray}
\label{udpropagator}
S_f(p)&=&ie^{-p^2_\perp l_f^2}\sum_{n=0}^{\infty}(-1)^n\frac{D_{f}(p)}{G(p)},\nonumber\\
D_{f}(p)&=&{2[p_0\gamma^0-p^3\gamma^3+m][\mathcal{P}_+L_n(2p^2_\perp l_f^2)-\mathcal{P}_-L_{n-1}(2p^2_\perp l_f^2)]+4\vec{\gamma}_\perp\cdot \vec{p}_\perp L_{n-1}^1(2p_\perp^2l_f^2)},\nonumber\\
G(p)&=&p_0^2-p_3^2-2n|q_fB|-m_f^2,
\end{eqnarray}
where $\mathcal{P}_\pm=\frac{1}{2}[1\pm i\gamma^1\gamma^2 s_\perp^f]$ is the projector. For u-quark $s_\perp^u=+1$, $l_u^2=\frac{3}{2}l^2$, and for d-quark $s_\perp^d=-1$, $l_d^2=3l^2$. 

\paragraph{Gap equation and constituent quark mass}
In the QCD vacuum, quark-antiquark pairs condense, leading to the generation of dynamical masses for quarks. Assuming isospin symmetry with $m_u = m_d = m_0$, the constituent quark mass $m_f$ for light quarks is determined by solving the gap equation $m_f~=~m_0+2G_S\text{Tr}S(x,x)$. Then, the gap equation is 
rotating to Euclidean space
\begin{eqnarray}
\label{gap}
m_f&=&m_0+8 G_S m_f N_c\sum_{f=u,d}\frac{|q_fB|}{4\pi}\sum_{n=0}^{\infty} \alpha_n\int_{-\infty}^\infty\frac{dp_3}{2\pi}\frac{1-2n_F(E_F)}{2E_f},
\end{eqnarray}
where $\alpha_n=1 $ for $n=0$, and $\alpha_n=2 $ for $n\geq1$ is the degree of degenerency of spin. $n_F=1/(e^{x/T}+1)$ is the Fermi Dirac distribution, the energy eigenvalue of quark in the magnetic field $E_f~=~\sqrt{2n|q_fB|+p_3^2+m_f^2}$. This NJL mean field gap equation reproduce the magnetic catalysis at $T=0.01$GeV as expected in other model analysis.

\paragraph{Meson quadratic kernel and polarization matrix}
We now turn to investigate the properties of mesons. To keep the lagrangian minimal, we focus on the $\pi^\pm$ and $\rho^\pm$ sector, and focus on the $s_z=0$ part of $\rho^\pm$. To avoid the coupling between $\rho^\pm_0$ and $\rho^\pm_3$, we restrict ourself to $q_z=0$. This is a consistent kinematic truncation for extracting pole energies of the lowest Landau level at zero longitudinal momentum. Restoring $q_z\neq 0$ would enlarge the kernel but not change the qualitative mixing mechanism. With these restrictions, we have the Hubbard–Stratonovich transformed Lagrangian
\begin{eqnarray}
\mathcal L_{\text{HS}}=\bar\psi\big(iD\!\!\!\!/ - m_f\big)\psi
-\bar\psi\Big[i\gamma_5\tau^+\pi^-+i\gamma_5\tau^-\pi^+
+\gamma_\mu\tau^+\rho^{-\mu}+\gamma_\mu\tau^-\rho^{+\mu}
+\cdots
\Big]\psi
-\frac{|\pi|^2}{4G_S}
+\frac{|\rho|^2}{4G_V}
+\cdots.
\end{eqnarray}
After integrating out quarks and expand to quadratic order, the bosonized effective action is
\begin{eqnarray}
\label{bosonizedaction}
S_{\rm eff}[\pi,\rho]=
\int d^4x\left(
\frac{\pi^+\pi^-}{4G_S}
-\frac{\rho^+_\mu\rho^{-\mu}}{4G_V}
\right)
-i\mathrm{Tr}\ln\Big[ S_f^{-1} - \hat\Sigma(\pi,\rho)\Big]
+\int d^4x\mathcal L_{\text{tree}},
\end{eqnarray}
with $S_f^{-1}=iD\!\!\!\!/-m_f$ in the background field, and $\hat\Sigma=i\gamma_5\tau^+\pi^-+i\gamma_5\tau^-\pi^++\gamma_\mu\tau^+\rho^{-\mu}+\gamma_\mu\tau^-\rho^{+\mu}$ the meson–quark vertices. Throughout this work we work at the RPA level and retain only quadratic terms in the meson fields. By expanding the fermion determinant to second order we have to the second order
\begin{eqnarray}
\label{actionexpansion}
-i\mathrm{Tr}\ln\big[S_f^{-1}-\hat\Sigma\big]
=
-i\mathrm{Tr}\ln S_f^{-1}
+\frac{i}{2}\mathrm{Tr}\big(S_f\hat\Sigma S_f\hat\Sigma\big)
+\cdots. 
\end{eqnarray}
So the quadratic meson action is
\begin{eqnarray}
\label{quadraaction}
S^{(2)}_{\rm eff}
=
\frac{1}{2} \int d^4x d^4y
\Phi^\dagger(x)
\Big[
\mathcal{G}^{-1}\delta(x-y)-\Pi(x,y)
\Big]
\Phi(y)
+S^{(2)}_{\rm tree},
\end{eqnarray}
where $\Phi=(\pi,\rho_3)^T$ in the reduced subspace. Here $\rho_3$ denotes the longitudinal $s_z=0$ component of the charged rho meson. $\mathcal G^{-1}$ is the “contact” inverse propagator from the HS terms, and $\Pi$ is the polarization matrix from the quark loop. $S^{(2)}_{\rm tree}=\int d^4x\mathcal L_{\text{tree}}$ comes from the tree level counterterm. In the Random Phase Approximation (RPA), the core ingredient is the polarization function, which is defined here by 
\begin{eqnarray}
\Pi(x,y)&=&\left(\begin{array}{cc}
\Pi_{\pi\pi} &  \Pi_{\pi\rho}^3 \\
\Pi_{\rho\pi}^3  &  \Pi_{\rho\rho}^{33} \\
\end{array}\right),\nonumber\\
\Pi_{\pi\pi}&=&-i\text{Tr}\big[(i\gamma^5\tau^-)S_f(r,r')(i\gamma^5\tau^+)S_f(r',r)\big],\nonumber\\
\Pi_{\rho\rho}^{33}&=&-i\text{Tr}\big[(\gamma^3\tau^-)S_f(r,r')(\gamma^3\tau^+)S_f(r',r)\big],\nonumber\\
\Pi_{\rho\pi}^3&=&-i\text{Tr}\big[(\gamma^3\tau^-)S_f(r,r')(i\gamma^5\tau^+)S_f(r',r)\big],\nonumber\\
\Pi_{\pi\rho}^3&=&-i\text{Tr}\big[(i\gamma^5\tau^-)S_f(r,r')(\gamma^3\tau^+)S_f(r',r)\big].
\end{eqnarray}
The off-diagonal component $\Pi_{\rho\pi}^3$ and $\Pi_{\pi\rho}^3$ are non-vanishing in presence of the external magnetic field. They introduce the mixing between $\pi^\pm$ and $\rho^\pm_{s_z=0}$ at loop level.  

\paragraph{Regularization scheme and model parameters}
In the calculation, we use the Pauli Villars regulation with three regulators $a_i=\{0,1,2,3\}$, $c_i=\{1,-3, 3,-1\}$. The model parameters are chosen to be $\Lambda=1.25$GeV, $G_S=3.1$GeV$^{-2}$, $G_V=5.019$GeV$^{-2}$ for all schemes except the bosonization. These parameters correspond to $m_\pi=0.133$GeV, $m_\rho=0.77$GeV. The summation of Landau level is performed to over 200 Landau levels, which is sufficient for the range of magnetic field considered.  

\paragraph{Weak-field loop-induced $\pi-\rho$ mixing and its role in the matching}
Before fixing the phenomenological mixing strength from the radiative decay $\rho^\pm\to\pi^\pm\gamma$, we first separate the part of the mixing vertex that is generated explicitly by the quark loop in the present NJL truncation. This can be done most transparently in the weak-field limit. Since the off-diagonal polarization functions $\Pi^3_{\rho\pi}$ and $\Pi^3_{\pi\rho}$ vanish in vacuum and are odd in the external magnetic field, their leading contribution is linear in $B$. We therefore expand the quark propagator to first order in the magnetic field and define the loop-induced mixing strength as the coefficient of the structure $iq_0 eB$.

This separation is useful for the matching procedure. The background-field polarization function and the vacuum $\rho\pi\gamma$ transition amplitude should not be regarded as two unrelated effects. At linear order in the external magnetic field, the magnetic background acts as one soft electromagnetic insertion on the quark line. The resulting off-diagonal $\pi-\rho$ polarization is therefore the soft external-field limit of the same quark-level triangle diagram that contributes to the vacuum radiative transition $\rho\to\pi\gamma$. In this sense, the weak-field calculation identifies the microscopic loop part of the effective $\rho\pi\gamma$ vertex within the present model.

For a quark of flavor $f$ and electric charge $q_f=Q_f e$ in a constant magnetic field along the $z$ direction, $\mathbf{B}=B\hat z$, the translationally invariant part of the propagator can be expanded as
\begin{equation}
S_f(k;B)=S_f^{(0)}(k)+S_f^{(1)}(k;B)+{\cal O}(B^2),
\end{equation}
where
\begin{equation}
S_f^{(0)}(k)= \frac{k\cdot \gamma+M}{k^2-M^2+i\epsilon}, \qquad
S_f^{(1)}(k;B)= iq_fB\, \frac{\gamma^1\gamma^2(k_\parallel\cdot \gamma_\parallel+M)}{(k^2-M^2+i\epsilon)^2}.
\end{equation}
Here $k^\mu_{\parallel}=(k^0,0,0,k^3)$ and $k^\mu_{\perp}=(0,k^1,k^2,0)$. Keeping only terms linear in $B$, the off-diagonal polarization function for the charged channel becomes
\begin{align}
\Pi^3_{\rho^+\pi^+}(q)
=&
-i\,2N_c
\int\frac{d^4k}{(2\pi)^4}
{\rm Tr}\big[
\gamma^3 S_u^{(0)}(k)i\gamma^5 S_d^{(1)}(p;B)
+\gamma^3 S_u^{(1)}(k;B)i\gamma^5 S_d^{(0)}(p)
\big]
\nonumber\\
\equiv& -i q_0 eB\, g_{\rho\pi}^{\rm loop,vac},
\end{align}
with $p=k+q$. After carrying out the frequency integral, the vacuum loop
coefficient can be written as
\begin{equation}
g_{\rho\pi}^{\rm loop,vac}
=
\frac{8}{3}N_c M
\int\frac{d^3k}{(2\pi)^3}
\left[
\frac{12E_f^2-q_0^2}
     {4E_f^3(q_0^2-4E_f^2)^2}
(1-2n_F(E_f))
-
\frac{n_F'(E_f)}
     {2E_f^2(q_0^2-4E_f^2)}
\right],
\end{equation}
where $E_f=\sqrt{\mathbf{k}^2+M^2}$. Evaluating this expression in vacuum with $M=0.4~{\rm GeV}$ and $q_0=m_\pi=0.133~{\rm GeV}$ gives
\begin{equation}
g_{\rho\pi}^{\rm loop,vac}=0.064556 .
\end{equation}

The numerical value of $g_{\rho\pi}^{\rm loop,vac}$ shows that the quark loop provides a sizable part of the effective $\rho\pi\gamma$ coupling, but it does not saturate the coupling inferred from the physical radiative decay. This is the reason why, in the following matching, we do not identify the weak-field loop result with the full vacuum transition strength. Instead, we treat $g_{\rho\pi}^{\rm loop,vac}$ as the explicitly resolved microscopic contribution in the present truncation, and determine the remaining local part by matching
the full effective vertex to the experimental decay width.

This prescription also fixes how the magnetic-field dependence is implemented in the later sections. The quantity $g_{\rho\pi}^{\rm loop}(B)$ is computed from the quark-loop polarization function and therefore carries the magnetic-field dependence generated by the microscopic NJL dynamics. The part not resolved by this loop calculation is represented by a local counterterm fixed in vacuum and kept as a constant effective contribution. Thus the mixing strength used below has the form $g_{\rho\pi}(B)
=g_{\rho\pi}^{\rm loop}(B)+g_{\rho\pi}^{\rm tree}$.

This should be understood as a minimal matching scheme: the weak-field polarization determines the loop component of the $\rho\pi\gamma$ vertex, while the physical decay width fixes the total vacuum normalization and hence the remaining local contribution.

\paragraph{Tree-level $\pi-\rho$ mixing operator and vacuum matching}
The weak-field calculation above determines the part of the $\rho\pi\gamma$ vertex that is explicitly generated by the quark loop in the present NJL truncation. To reproduce the physical radiative transition, however,
one has to match the total effective vertex to the experimental $\rho^\pm\to\pi^\pm\gamma$ decay width. We implement this matching by adding the lowest order gauge-invariant local operator that is linear in the electromagnetic field and connects the charged pion to the longitudinal charged rho mode,
\begin{equation}
{\cal L}_{\rm tree} =\kappa_{\rm tree}
\left(\rho^+_\mu \widetilde F^{\mu\nu}D_\nu\pi^-+\rho^-_\mu \widetilde F^{\mu\nu}D_\nu\pi^+\right),
\end{equation}
where $\widetilde F^{\mu\nu}=\epsilon^{\mu\nu\alpha\beta}F_{\alpha\beta}/2$ and
$\kappa_{\rm tree}$ has mass dimension $-1$. In a static magnetic field along
the $z$ direction, this operator reduces in the $q_z=0$, $\rho_3$ sector to a term proportional to $\rho_3^\pm\partial_0\pi^\mp$. Thus it contributes directly to the off-diagonal
quadratic kernel as
\begin{equation}
\label{tree_kernel}
K^{\rm tree}_{\rho\pi}(q_0;B)=+i q_0\,\kappa_{\rm tree}B,
\qquad
K^{\rm tree}_{\pi\rho}(q_0;B)=-i q_0\,\kappa_{\rm tree}B .
\end{equation}

The role of this operator is purely matching. It should not be interpreted as an independent mechanism unrelated to the loop-induced polarization discussed above. Rather, the full low-energy $\rho\pi\gamma$ vertex is decomposed into an explicitly calculated quark-loop contribution and a remaining local contribution that is not resolved within the present truncation. The latter is represented by ${\cal L}_{\rm tree}$ and fixed by vacuum data.

To determine the normalization, consider the radiative decay $\rho(P,\epsilon_\rho)\rightarrow \pi(p)+\gamma(q,\epsilon_\gamma)$, with $P=p+q$. For an outgoing photon, the dual field strength gives $\widetilde F^{\mu\nu}\rightarrow i\epsilon^{\mu\nu\alpha\beta}q_\alpha\epsilon_{\gamma,\beta}$, while the derivative acting on the pion gives $D_\nu\pi\rightarrow -ip_\nu\pi$. The corresponding amplitude can be written as
\begin{equation}
i{\cal M}
=
i\kappa_{\rm exp}
\epsilon^{\mu\nu\alpha\beta}
\epsilon_{\rho,\mu}
\epsilon_{\gamma,\beta}
p_\nu q_\alpha .
\end{equation}
After summing over final polarizations and averaging over the three rho
polarizations, one obtains
\begin{equation}
\overline{|{\cal M}|^2}
=
\frac{2}{3}\kappa_{\rm exp}^2(P\cdot q)^2 .
\end{equation}
The two-body decay width is therefore
\begin{equation}
\Gamma(\rho\rightarrow\pi\gamma)
=
\frac{\kappa_{\rm exp}^2}{96\pi m_\rho^3}
\left(m_\rho^2-m_\pi^2\right)^3 .
\end{equation}
Using the experimental partial width
$\Gamma_{\rm exp}(\rho^\pm\rightarrow\pi^\pm\gamma)\simeq 0.068~{\rm MeV}$,
we obtain $\kappa_{\rm exp}\simeq 0.222~{\rm GeV}^{-1}$.

For use in the quadratic meson kernel, the coupling must be expressed in terms
of the normalization convention of the composite pion and rho fields. In the
pole-based schemes considered below, namely the rest-mass, direct-determinant,
and near-pole extractions, we define the vacuum wave-function normalization
factors from the slopes of the diagonal polarization functions at the physical
poles,
\begin{equation}
Z_\pi^{\rm vac}
=
\left.
\frac{\partial \Pi_{\pi\pi}(q_0)}
     {\partial q_0^2}
\right|_{q_0^2=m_\pi^2},
\qquad
Z_\rho^{\rm vac}
=
\left.
\frac{\partial \Pi_{\rho\rho}(q_0)}
     {\partial q_0^2}
\right|_{q_0^2=m_\rho^2}.
\end{equation}
With this convention, the inverse pole residue is $Z_\phi^{\rm vac}$, and the
canonically normalized mixing strength contains the factor
$(Z_\pi^{\rm vac}Z_\rho^{\rm vac})^{-1/2}$. The vacuum polarization functions entering this definition are
\begin{align}
\Pi_{\pi\pi}(q_0) &= -i\int\frac{d^4k}{(2\pi)^4}{\rm Tr}\left[i\gamma_5\tau^- S(k)i\gamma_5\tau^+ S(p)\right]=
8N_c
\int\frac{k^2dk}{2\pi^2}
\frac{2E_f(1-2n_F(E_f))}
     {4E_f^2-q_0^2},
\\
\Pi_{\rho\rho}(q_0)
&=
-i\int\frac{d^4k}{(2\pi)^4}
{\rm Tr}
\left[
\gamma^3\tau^- S(k)
\gamma^3\tau^+ S(p)
\right]
=
8N_c
\int\frac{k^2dk}{2\pi^2}
\left(
\frac{4E_f}{3}
+
\frac{2M^2}{3E_f}
\right)
\frac{1-2n_F(E_f)}
     {4E_f^2-q_0^2}.
\end{align}
Numerically, for the parameter set used in the pole-based calculations, this
gives $Z_\pi^{\rm vac}=0.109$ and $Z_\rho^{\rm vac}=0.413$.

The full vacuum transition strength inferred from the decay width must then be
reproduced by the sum of the loop contribution and the local remainder. With
$\kappa_{\rm exp}=e\,g_{\rho\pi}^{\rm exp}$, the matching condition is
\begin{equation}
\frac{g_{\rho\pi}^{\rm loop,vac}+g_{\rho\pi}^{\rm tree}}{\sqrt{Z_\pi^{\rm vac}Z_\rho^{\rm vac}}} = \frac{\kappa_{\rm exp}}{e} \simeq 0.73337~{\rm GeV}^{-1}.
\end{equation}
Using the weak-field loop result $g_{\rho\pi}^{\rm loop,vac}=0.064556$, this fixes
\begin{equation}
g_{\rho\pi}^{\rm tree} = 0.09104~{\rm GeV}^{-1}.
\end{equation}
The total mixing coefficient used in the pole-based extraction schemes is
therefore
\begin{equation}
g_{\rho\pi}(B) = g_{\rho\pi}^{\rm loop}(B) + g_{\rho\pi}^{\rm tree}.
\end{equation}
Here $g_{\rho\pi}^{\rm loop}(B)$ is calculated from the off-diagonal quark
polarization function at finite magnetic field, while
$g_{\rho\pi}^{\rm tree}$ is kept as the constant local remainder fixed by the
vacuum radiative decay. This prescription keeps the magnetic-field dependence
that is explicitly generated by the microscopic loop calculation, while ensuring
that the total vacuum $\rho\pi\gamma$ coupling is normalized to experiment.

In the local bosonized scheme, the same matching logic is used, but the
normalization factors are not the pole residues above. They are instead obtained
from the derivative expansion around vanishing external momentum. Consequently,
the numerical value of the local tree-level remainder differs from that in the
pole-based schemes. This point will be discussed separately in the local
extraction section.

The ingredients introduced above — the NJL model in a magnetic background, the Landau-level quark propagator, the quadratic kernel in the $(\pi^\pm,\rho^\pm_{s_z=0})$ sector, and the matched local mixing operator — provide the common microscopic input for all four extraction schemes. What differs in the following sections is not the underlying dynamics, but how this same kernel is projected onto physical masses or energies. We begin with the most direct extension of the vacuum pole definition, namely the rest-mass construction.


\section{Rest-Mass Extraction (RPA Pole at Zero Momentum)}
\label{sec_rest}
We start from the most direct extension of the familiar vacuum pole definition. In NJL-type approaches, the meson mass in vacuum is usually determined from the pole of the random-phase-approximation (RPA) propagator at zero spatial momentum. In a magnetic field, the simplest generalization is to apply the same prescription to the coupled $\pi-\rho$ kernel and define the mass through
\begin{eqnarray}
\text{det}~\mathcal{K}(q_0,\vec{q}=0;B)=0
\end{eqnarray}
This construction answers the question: what is the energy required to create the composite excitation at rest?

This definition provides a natural baseline, because it stays as close as possible to the usual vacuum RPA pole construction. At the same time, it is also the least adapted to the kinematics of a charged excitation in a magnetic field. The magnetic field enters through the quark-loop polarization functions and the mixing kernel, but no explicit Landau-level structure is imposed on the external meson state. The resulting quantity is therefore a rest-frame pole of the composite correlator, rather than the energy of a propagating charged Landau eigenmode.

For this reason, the rest-mass scheme is best viewed as a reference point for the later analysis. It isolates how much of the magnetic-field effect can already be captured by the composite pole at $\vec{q}=0$, before one incorporates Landau-level projection or quasiparticle normalization more explicitly.

In the charged $\pi^\pm-\rho^\pm_{s_z=0}$ system, the quadratic kernel takes the form
\begin{eqnarray}
\label{RPAkernal}
\mathcal{K}(q_0,B)&=&\left(\begin{array}{cc}
\frac{1}{2G_S}-\Pi_{\pi\pi}(q_0,B)  &  -{\Pi}^3_{\pi\rho}(q_0,B)+\mathcal{K}_{\pi\rho}^{\text{tree}} \\
-{\Pi}^3_{\rho\pi}(q_0,B)+\mathcal{K}_{\rho\pi}^{\text{tree}}   &    \frac{1}{2G_V}-\Pi_{\rho\rho}(q_0,B)\\
\end{array}\right).
\end{eqnarray}
The tree counterterm is the truncated to $q_z=0$, $\rho\rightarrow\rho_3$ sector, and contribute to the kernel as \eqref{tree_kernel}. In the meson rest frame $q=(q_0,0)$, the polarization function are given by
\begin{eqnarray}
\Pi_{\pi\pi}(q_0,B)&=&-\frac{4N_ceB}{3\pi}\sum_{m,n=0}^{\infty}\int\frac{dk_3}{(2\pi)}(-1)^{n+m}\Big(H_1(n,m)I_3(q_0)+[(k_3^2+m^2)H_1(n,m)-\frac{8eB}{3}H_2(n,m)]I_2(q_0)\Big),\nonumber\\\Pi_{\rho\rho}(q_0,B)&=&-\frac{4 N_c eB}{3\pi}\sum_{m,n=0}^{\infty}\int\frac{dk_3}{(2\pi)}(-1)^{n+m}\Big(H_1(n,m)I_3(q_0)-[(k_3^2-m_f^2)H_1(n,m)+\frac{8eB}{3} H_2(n,m)]I_2(q_0)\Big),\nonumber\\
\Pi_{\rho\pi}^{3}(q_0,B)&=&-i \frac{4N_c eB}{3\pi} m_f q_0\sum_{m,n=0}^\infty\int\frac{dk_3}{(2\pi)}(-1)^{n+m}H_3(n,m)I_2(q_0)\equiv- iq_0\, eB\, g_{\rho\pi}^\text{loop}(q_0,B),
\end{eqnarray}
where $n$ and $m$ are Landau levels in the u- and d-quark propagator respectively, $I_2$ and $I_3$ are the the functions encoding the Matsubara sum are defined in \eqref{I2I3q0} in Appendix.\ref{App_matsubara}, and $H_i$ are integrals over Laguerre polynomials generated by the integral over perpendicular momentum. The definitions and analytical expressions of $H_i$ are included in the Appendix.\ref{App_H}. The other off-diagonal component is given by $\Pi_{\pi\rho}^{3}=\Pi_{\rho\pi}^{3*}=iq_0\, eB\, g_{\rho\pi}^\text{loop}(q_0,B)$. With $g_{\rho\pi}^\text{loop}(q_0,B)$ describes the dressing of the loop-induced mixing 
\begin{eqnarray}
g_{\rho\pi}^\text{loop}(q_0,B)&=& \frac{4N_c }{3\pi} m_f \sum_{m,n=0}^\infty\int\frac{dk_3}{(2\pi)}(-1)^{n+m}H_3(n,m)I_2(q_0). 
\end{eqnarray}
Hence the off-diagonal component of the kernel can be rewritten as $-iq_0H(q_0, B)\equiv -{\Pi}^3_{\pi\rho}+\mathcal{K}_{\pi\rho}^{\text{tree}}$ and $iq_0H(q_0, B)\equiv -{\Pi}^3_{\rho\pi}+\mathcal{K}_{\rho\pi}^{\text{tree}}$, with $H(B)=(g_{\rho\pi}^\text{loop}+g^\text{tree}_{\rho\pi}) eB$ in the kernel denotes the combined loop-induced and tree-level mixing coefficient, with $g^\text{tree}_{\rho\pi}$ determined in Sec.\ref{com_input}

The rest-mass spectrum is obtained by solving
\begin{eqnarray}
\label{rest_eigen}
\text{det}\mathcal{K}(q_0,B)=\Big(\frac{1}{2G_S}-\Pi_{\pi\pi}\Big)\Big(\frac{1}{2G_V}-\Pi_{\rho\rho}\Big)-q_0^2H(B)^2=0,
\end{eqnarray}
for real $q_0=m_\pm(B)$. In the absence of mixing, the pion and longitudinal rho masses are determined independently from the diagonal components. When mixing is present, the eigenvalues of the coupled system define two rest-frame pole masses. 

This scheme preserves the familiar RPA pole construction and therefore serves as the natural baseline for comparison. Its limitation is equally clear: although the magnetic field modifies the internal quark dynamics, the external charged meson is still treated as if it were created in a plane-wave rest frame. As a result, the extracted pole does not yet correspond to the energy of a Landau-quantized charged mode.

\begin{figure}[H]
\centering
\includegraphics[width=0.4\textwidth]{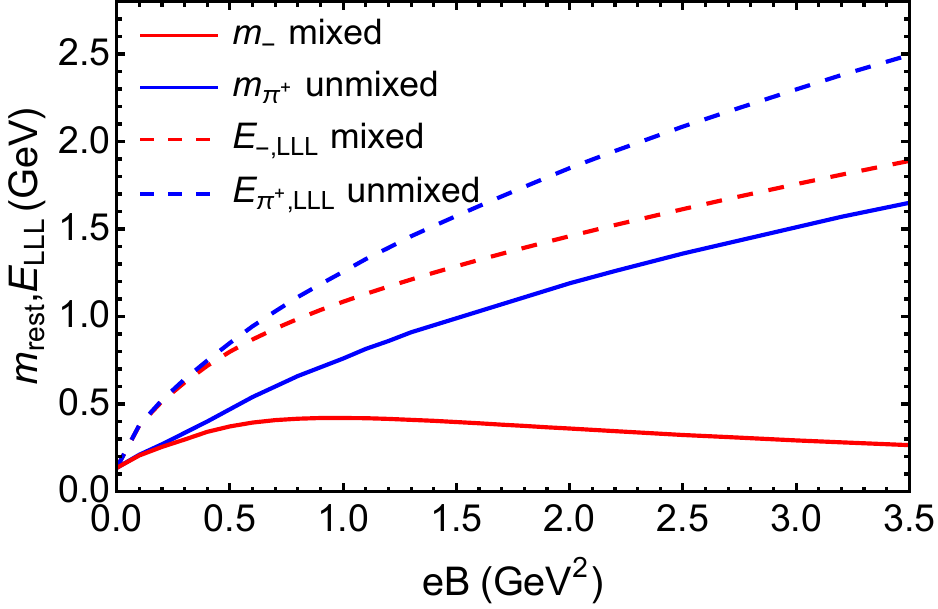}
\caption{
Rest-mass extraction. Solid lines show the unmixed charged-pion rest mass and the lower mixed rest-mass eigenvalue obtained from the RPA pole condition at $\mathbf{q}=0$. Dashed lines show the corresponding reconstructed lowest-Landau-level energies, $E_{\rm LLL}=(m_{\rm rest}^2+eB)^{1/2}$. Although the mixed rest mass decreases at large magnetic field, the reconstructed LLL energy remains monotonic.}
\label{rest_1}
\end{figure}

In Fig.\ref{rest_1} we present the magnetic field dependence of unmixed charged pion rest-mass $m_\pi$ as well as the lower rest-mass $m_-(B)$ of the mixed system. We also present the energy of the lowest landau level by adding the corresponding kinetic energy $E_\phi=(m_\text{rest}^2+eB)^{1/2}$. Although the lower eigenvalue of rest-mass in the mixed system exhibits a non-monotonic behavior with magnetic field over the range considered, its corresponding LLL energy $E_{LLL}$ is monotonic increasing with magnetic field. The inclusion of $\pi-\rho$ mixing in such kinetic slice produces only a modest quantitative shift relative to the unmixed case.

In the rest-mass scheme, the pole is determined in the meson rest frame, and the lowest Landau level (LLL) energy is subsequently reconstructed through $E_\phi=(m_\text{rest}^2+eB)^{1/2}$. As long as the extracted rest mass squared remains positive, the LLL energy is bounded from below by $\sqrt{eB}$. However, lattice simulations show that at sufficiently large magnetic field the lowest eigenmode falls below $\sqrt{eB}$, implying that the effective mass parameter entering the dispersion relation cannot remain strictly positive. Within the rest-mass prescription adopted here, the mixing effect modifies the rest mass only moderately and does not drive it negative. As a result, the LLL energy in this scheme never drops below the $\sqrt{eB}$ baseline and therefore cannot reproduce the non-monotonic behavior observed in lattice QCD.

The rest-mass definition therefore provides the simplest operational notion of mass in a magnetic background, but it does so by keeping the pole prescription in the vacuum style. Its failure to reproduce the lattice turnover is not merely quantitative; it reflects a structural limitation of describing a charged excitation in a magnetic field without imposing Landau-level kinematics on the external meson state.

\section{Local extraction}
\label{sec_local}
The failure of the rest-mass construction shows that, in a magnetic field, the relevant charged excitation cannot in general be described by extracting a vacuum-style pole at $\vec{q}=0$ and then reconstructing the lowest Landau level energy afterward. A natural next step is therefore to reorganize the same microscopic kernel into a local mesonic effective theory, in which the external magnetic field enters directly through covariant derivatives acting on charged meson fields.

This leads to the local-extraction, or bosonization, scheme. Instead of determining the spectrum directly from the full nonlocal kernel, we bosonize the underlying quark model and perform a derivative expansion of the effective action. The result is a local quadratic mesonic theory in which the magnetic-field dependence is encoded in effective kinetic coefficients, mass parameters, and mixing terms. This construction answers a different question from the rest-mass scheme: what effective mass and mixing parameters govern the charged meson system in a local effective theory?

A key feature of this approach is that the nontrivial Schwinger phases appearing in the quark polarization functions are automatically reorganized into covariant derivatives of the effective charged fields \cite{Wang:2017vtn}. In this sense, the bosonized description incorporates the gauge structure of the external magnetic field already at the level of the meson effective action. At the same time, however, the expansion is performed around small external momentum, so the resulting coefficients are local quantities rather than pole residues of the physical charged mode. The local scheme therefore provides an effective field theory reinterpretation of the same microscopic dynamics, but not yet a direct extraction of the Landau-projected quasiparticle pole.

We now derive the local quadratic action in the coupled $\pi^\pm-\rho^\pm_{s_z=0}$ sector and identify the corresponding magnetic-field-dependent parameters entering the local mass extraction.

\paragraph{Local expansion of the pion sector.}
We illustrate the procedure by first considering the diagonal pion sector. The $\pi^+\pi^-$ component in the quadratic action in \eqref{quadraaction} comes from the corresponding term in $\frac{i}{2}\mathrm{Tr}\big(S_f\hat\Sigma S_f\hat\Sigma\big)$ \eqref{actionexpansion}
\begin{eqnarray}
S^{(2)}_{\pi^+\pi^-}
&=&\frac{i}{2}\times 2N_c\text{Tr}\int d^4x d^4y e^{i\Phi_u(x,y)+i\Phi_d(y,x)} (i\gamma^5)\widetilde{S}_u(x-y)(i\gamma^5)\widetilde{S}_d(y-x)\pi^+(x)\pi^-(y)
\end{eqnarray}
together with the contact term $\frac{1}{2} \int d^4x \frac{1}{2G_S}\pi^-\pi^+$, where $\widetilde{S}_f(x-y)$ is the translation invariant part of the fermion propagator, while $e^{i\Phi_u(x,y)+i\Phi_d(y,x)} $ is the Schwinger phase. The combined Schwinger phase satisfies
\begin{equation}
\Phi_u(x,y)+\Phi_d(y,x) = Q_u\int_y^xA^\mu(x')d x'_\mu+Q_d\int_x^yA^\mu(x')d x'_\mu =\Phi_{\pi_-}(y,x)=\Phi_{\pi_+}(x,y).
\end{equation}
which corresponds to a Wilson line associated with the finite displacement of the charged pion in the background magnetic field. To obtain a local effective action, both the pion field and the phase factor must be expanded consistently around x. Expanding to second order in $(y-x)$, we write
\begin{eqnarray}
\label{exp2}
\pi_-(y) &=& \pi_-(x)+(y-x)^\mu\partial_\mu \pi_-(x) +\frac{1}{2}(y-x)^\mu (y-x)^\nu \partial_\mu\partial_\nu \pi_-(x)+\cdots,\nonumber\\
e^{i\Phi_{\pi_-}(y,x)} &=& 1-ie A^\mu(x) (y-x)_\mu- i e \frac{1}{2}\frac{\partial A^\mu(x)}{\partial x^\nu} (y-x)_\mu  (y-x)_\nu+\frac{(-i e)^2}{2}\left(A^\mu(x) (y-x)_\mu\right)^2+\cdots,
\end{eqnarray}
Combining these expansions, the polarization contribution can be rewritten in terms of covariant derivatives $D^-_\mu = \partial_\mu - ie A_\mu(x)$.
\begin{eqnarray}
S^{(2)}_{\pi^+\pi^-} &=& i N_c\int_{x,y}\text{Tr}\widetilde{S}_u(x-y)i\gamma^5\widetilde{S}_d(y-x)i\gamma^5\nonumber\\
&&\times\Big[\pi_-(x)+(y-x)^\mu D^-_\mu\pi_-(x)+\frac{1}{2}(y-x)^\mu (y-x)^\nu D^-_\mu  D^-_\nu\pi_-(x)\Big]\pi_+(x)
\end{eqnarray}
plus the contact term. Likewise there is also a term $S^{(2)}_{\pi^-\pi^+}$. The two terms linear in $(y-x)^\mu$ in $S^{(2)}_{\pi^+\pi^-}$ and $S^{(2)}_{\pi^-\pi^+}$ can be summed up to give a surface term which vanishes in the sense of coordinate integration. Collecting kinetic and mass terms, the pion contribution to the effective action becomes
\begin{eqnarray}
S^{(2)}_{\pi\pi}&=&\frac{1}{2}\int_x\Big[{Z}^{\mu\mu}_\pi\Big(\big|D_\mu\pi_+\big|^2+\big|D_\mu\pi_-\big|^2\Big)-\frac{1}{2}(\frac{1}{2G_S}-m_\pi^2)\Big(|\pi_+|^2+|\pi_-|^2\Big)\Big],
\end{eqnarray}
where the wave function renormalization in the above expansion are defined by the following
\begin{equation}
{Z}^{\mu\mu}_\pi=-i N_c\int d^4z\int \frac{d^4p}{(2\pi)^4}\frac{d^4k}{(2\pi)^4}\text{Tr}\Big[\widetilde{S}_u(p)(i\gamma^5)\widetilde{S}_d(k)(i\gamma^5)\Big]e^{i(k-p)z}z_\mu z_\mu.
\end{equation}
For later convenience, we restrict ourself to the condition $q_z=0$, the anisotropic form 
\begin{eqnarray}
\label{pionaction}
S^{(2)}_{\pi\pi}&=&\frac{1}{2}\int_x\Big[
{Z}^\parallel_\pi\big(\big|D_0\pi_+\big|^2+\big|D_0\pi_-\big|^2\big)
-{Z}^\perp_\pi\big(\big|D_\perp\pi_+\big|^2+\big|D_\perp\pi_-\big|^2\big)
-(\frac{1}{2G_S}-m_\pi^2)\big(|\pi_+|^2+|\pi_-|^2\big)\Big]
\end{eqnarray}
is obtained, with ${Z}^\parallel_\pi={Z}^{00}_\pi$ and ${Z}^\perp_\pi={Z}^{11}_\pi={Z}^{22}_\pi$
\begin{eqnarray}
\label{piparameter}
m_\pi^2&=&-2i N_c\int\frac{d^4p}{(2\pi)^4} \text{Tr} \Big[\widetilde{S}_u(p)(i\gamma^5)\widetilde{S}_d(p)(i\gamma^5)\Big], \nonumber\\
{Z}^\parallel_\pi&=&-i N_c\int \frac{d^4p}{(2\pi)^4}\text{Tr}\Big[(\partial_0^p\partial_0^p\widetilde{S}_u(p))(i\gamma^5)\widetilde{S}_d(p)(i\gamma^5)\Big], \nonumber\\
{Z}^\perp_\pi&=&-i N_c\int \frac{d^4p}{(2\pi)^4}\text{Tr}\Big[(\partial_1^p\partial_1^p\widetilde{S}_u(p))(i\gamma^5)\widetilde{S}_d(p)(i\gamma^5)\Big].
\end{eqnarray}

\paragraph{Local expansion of the rho sector.} The local expansion of the rho sector proceeds analogously but yields a tensor structure characteristic of a massive vector field. Before local extraction, we project onto the physical longitudinal polarization that can mix with the pion.

At $q_z=0$, the transversality condition $q\cdot\rho=0$ implies $\rho_0=0$, leaving only the longitudinal $s_z=0$ component $\rho_3$. Applying this projection to the Proca kinetic term in a magnetic field, the quadratic action reduces to

\begin{eqnarray}
S_{\rho\rho}^{(2)}=\frac{1}{2}\int_x\Big[
Z_{\rho}^{\parallel}\big(|D_0\rho_3^+|^2+|D_0\rho_3^-|^2\big)
-Z_{\rho}^{\perp}\big(|D_\perp\rho_3^+|^2+|D_\perp\rho_3^-|^2\big)
-(\frac{1}{2G_V}-m_\rho^2)\big(|\rho_3^+|^2+|\rho_3^-|^2\big)\Big],
\end{eqnarray}
with the coefficients derived in the same way as in the pion sector, 
\begin{eqnarray}
\label{rhoparameter}
m_\rho^2&=&-2i N_c\int\frac{d^4p}{(2\pi)^4} \text{Tr} \Big[\widetilde{S}_u(p)\gamma^3\widetilde{S}_d(p)\gamma^3\Big],\nonumber\\
Z_{\rho}^{\parallel}&=&-i N_c\int \frac{d^4p}{(2\pi)^4}\text{Tr}\Big[(\partial_0^p\partial_0^p\widetilde{S}_u(p))\gamma^3\widetilde{S}_d(p)\gamma^3\Big],\nonumber\\
Z_{\rho}^{\perp}&=&-i N_c\int \frac{d^4p}{(2\pi)^4}\text{Tr}\Big[(\partial_1^p\partial_1^p\widetilde{S}_u(p))\gamma^3\widetilde{S}_d(p)\gamma^3\Big].
\end{eqnarray}

\paragraph{Local expansion of the mixing term.} Finally, we consider the off-diagonal mixing. Restricting to the charged sector and the longitudinal rho component, the quadratic mixing contribution contains two pairs of terms 
\begin{eqnarray}
S_{\rho^+\pi^-}^{(2)}+S_{\rho^-\pi^+}^{(2)}
&=&+i N_c \text{Tr}\int d^4x d^4y S_u(x,y)\gamma^\mu S_d(y,x) (i\gamma^5) \rho^+_\mu(x)\pi^-(y)\nonumber\\
&&+i N_c \text{Tr}\int d^4x d^4y S_d(x,y)\gamma^\mu S_u(y,x)(i\gamma^5) \rho^-_\mu(x)\pi^+(y),\nonumber\\
S_{\pi^-\rho^+}^{(2)}+S_{\pi^+\rho^-}^{(2)}
&=&+i N_c \text{Tr}\int d^4x d^4y S_d(x,y)(i\gamma^5) S_u(y,x) \gamma^\mu \pi^-(x)\rho^+_\mu(y)\nonumber\\
&&+i N_c \text{Tr}\int d^4x d^4y S_u(x,y)(i\gamma^5) S_d(y,x)\gamma^\mu \pi^+(x)\rho^-_\mu(y).
\end{eqnarray}
Performing the same local expansion as above for the $\pi^\pm(y)$ and $\rho^\pm_\mu(y)$ fields, summing all the mixing terms above, and integrating by parts to turn the terms like $\pi_-D_\alpha\rho^+_\mu$ into $\rho^+_\mu D_\alpha \pi_-$, we have the mixing term in the gauge-invariant form up to a surface term
\begin{eqnarray}
S_{\rho\pi}^{(2)}
&=&\int d^4x\Big(H^{\mu\alpha}_{\rho\pi}\rho^+_\mu(x) D^-_\alpha\pi_-(x)+h.c.\Big)
\end{eqnarray}
in the external magnetic field, only $H^{30}_{\rho\pi}\equiv H_{\rho\pi}$ survives. Explicitly, 
\begin{eqnarray}
H_{\rho\pi}=+2N_c\int \frac{d^4p}{(2\pi)^4}\text{Tr}[(\partial_0^p\widetilde{S}_u(p)) \gamma^3 \widetilde{S}_d(p)(i\gamma^5)].
\end{eqnarray}
which is real and unique once Hermitian conjugation is accounted for. 

\paragraph{Summary of local bosonic action.} Collecting all terms, the local quadratic action in the $\pi^\pm-\rho^\pm_{s_z=0}$ subspace at $q_z=0$ reads
\begin{equation}
\label{bosonaction}
\begin{split}
S^{(2)}_{\rm eff}=\int d^4x\Big\{
&\frac{1}{2}{Z}^\parallel_\pi\big(|D_0\pi^+|^2+|D_0\pi^-|^2\big)
-\frac{1}{2}{Z}^\perp_\pi\big(|D_\perp\pi^+|^2+|D_\perp\pi^-|^2\big)
-\frac{1}{2}(\frac{1}{2G_S}-m_\pi^2)\big(|\pi^+|^2+|\pi^-|^2\big)\\
+&\frac{1}{2}Z_{\rho}^{\parallel}\big(|D_0\rho_3^+|^2+|D_0\rho_3^-|^2\big)
\,-\frac{1}{2}Z_{\rho}^{\perp}\big(|D_\perp\rho_3^+|^2\,+|D_\perp\rho_3^-|^2\big)
-\frac{1}{2}(\frac{1}{2G_V}-m_\rho^2)\,\big(|\rho_3^+|^2+|\rho_3^-|^2\big)\\
+&\big(H_{\rho\pi}+\kappa_\text{tree}\tilde F^{30}\big)\big( \rho^+_3 D_0\pi^-+ \rho^-_3 D_0\pi^+\big)
\Big\}, 
\end{split}
\end{equation}
where we have also added the tree level term in the mixing term. This serves as the starting point for the local mass extraction discussed below. After explicitly carrying out the trace and the momentum integral, 
the explicit mass term, loop mixing term and wave function renormalization parameters in the above bosonized Lagrangian \eqref{bosonaction} are
\begin{eqnarray}
\label{parameters}
m_\pi^2&=&-N_c\frac{4eB}{3\pi}\int\frac{dp_3}{2\pi}\sum_{n,m}(-1)^{m+n}\Big[H_1(n,m)I_{0d}^l-2|q_uB| (n H_1(n,m)+2 H_2(n,m))I_1^l\Big],\nonumber\\
m_\rho^2&=&-N_c\frac{4eB}{3\pi}\int\frac{dp_3}{2\pi}\sum_{n,m}(-1)^{m+n}\Big[H_1(n,m)I_{0d}^l-\Big((2n|q_uB|+2p_3^2) H_1(n,m)+4 |q_uB| H_2(n,m)\Big)I_1^l\Big],\nonumber\\
H_{\rho\pi}&=&N_cm_f\frac{4eB}{3\pi}\int\frac{dp_3}{2\pi}\sum_{n,m}(-1)^{m+n}H_3(n,m)I_1^l~\equiv~g^{\text{loop}}_{\rho\pi} eB,\nonumber\\
Z_\pi^\parallel&=&-N_c\frac{4eB}{3\pi}\int\frac{dp_3}{2\pi}\sum_{n,m}(-1)^{m+n}\Big[W_1H_1(n,m)+4 |q_uB|W_2 H_2(n,m) \Big],\nonumber\\
Z_\rho^\parallel&=&-N_c\frac{4eB}{3\pi}\int\frac{dp_3}{2\pi}\sum_{n,m}(-1)^{m+n}\Big[W_3H_1(n,m)+4 |q_uB|W_2 H_2(n,m) \Big],\nonumber\\
Z_\pi^\perp&=&N_c\frac{1}{\pi}\int\frac{dp_3}{2\pi}\sum_{n,m}(-1)^{m+n}\Big[H_4(n,m)I_{0d}^l-2|q_uB| (n H_4(n,m)+2 H_5(n,m))I_1^l\Big],\nonumber\\
Z_\rho^\perp&=&N_c\frac{1}{\pi}\int\frac{dp_3}{2\pi}\sum_{n,m}(-1)^{m+n}\Big[H_4(n,m)I_{0d}^l-\Big((2n|q_uB|+2p_3^2) H_4(n,m)+4 |q_uB| H_5(n,m)\Big)I_1^l\Big],
\end{eqnarray}
where $W_i$ presents the combinations defined by $W_1=-8n |q_uB| E_u^2 I_3^l+(2E_u^2+6n |q_uB|)I_2^l-I_1^l$, $W_2=3I_2^l-4E_u^2 I_3^l$, $W_3=-8 E_u^2 (n |q_uB|+p_3^2)I_3^l+(2 E_u^2+6n |q_uB|+6 p_3^2)I_2^l-I_1^l$. $H_i$ are integrals over Laguerre polynomials generated by the integral over perpendicular momentum. The definitions and analytical expressions of $H_i$ are included in the appendix.\ref{App_H}. $I_i^l$ are the Matsubara sum in loop functions, with the analytical expressions also presented in the appendix.\ref{App_matsubara}.

In the block that mixes $\Phi_+=(\pi_+, \rho_3^-)^T$, the quadratic Lagrangian can be written as $L^{(2)}_+=\frac{1}{2}\Phi_+^\dag\mathcal{K}_\text{loc}\Phi_+$ according to \eqref{bosonaction}, with the matrix $\mathcal{K}_\text{loc}$
\begin{eqnarray}
\mathcal{K}_\text{loc}=\left(\begin{array}{cc}
Z_\pi^\parallel q_0^2- Z_\pi^\perp(2n+1)|eB|-m_{\pi,\text{eff}}^2 &  -i q_0 H_\text{tot}(B)  \\
+i  q_0 H_\text{tot}(B)   &  Z_\rho^\parallel q_0^2- Z_\rho^\perp(2n+1)|eB|-m_{\rho,\text{eff}}^2    \\
\end{array}\right),
\end{eqnarray}
the mass terms are $m_{\pi,\text{eff}}^2=\frac{1}{2G_S}-m_\pi^2$ and $m_{\rho,\text{eff}}^2=\frac{1}{2G_V}-m_\rho^2$. The off-diagonal term $H_\text{tot}(B)=(g^{\text{tree}}_{\rho\pi}+g^{\text{loop}}_{\rho\pi}) eB$ contains both tree level and loop order mixing terms. The eigenmass corresponds to the lowest Landau level is obtained through the pole condition
\begin{eqnarray}
\text{det}\,\mathcal{K}_\text{loc}=0,
\end{eqnarray}
with the Landau level $n=0$, this generates two eigenmodes, with energies 
\begin{eqnarray}
\label{bosonization_eigen}
E_\pm=\frac{1}{\sqrt{2}}\sqrt{h^2+\tilde{E}_{\pi }^2+\tilde{E}_{\rho }^2\pm\sqrt{\Big(h^2+\tilde{E}_{\pi }^2+\tilde{E}_{\rho }^2\Big)^2-4\tilde{E}_{\pi }^2\tilde{E}_{\rho }^2}}.
\end{eqnarray}
where $\tilde{E}_{\pi }^2=(Z_\pi^\perp |eB|+m_{\pi,\text{eff}}^2)/ Z_{\pi }^\parallel$,  $\tilde{E}_{\rho }^2=(Z_\rho^\perp |eB|+m_{\rho,\text{eff}}^2)/ Z_{\rho}^\parallel$ and $h=(g^{\text{tree}}_{\rho\pi}+g^{\text{loop}}_{\rho\pi}) eB/\sqrt{ Z_{\pi }^\parallel Z_{\rho }^\parallel}$.

Before turning to the numerical results, we briefly comment on the vacuum matching in the local scheme. Since the bosonized effective action is normalized through a derivative expansion at vanishing external momentum, the tree-level mixing parameter entering the local Lagrangian must be matched using the corresponding local normalization factors, rather than the pole residues adopted in the pole-based schemes. The weak field limit of the loop mixing in the local expansion yields $g^\text{loop,vac}_{\rho\pi}=0.0633257$. Together with the tree level term, they would produce the partial decay width of rho meson in vacuum
\begin{eqnarray}
\label{local_match}
(g^\text{loop,vac}_{\rho\pi}+g^\text{tree}_{\rho\pi})\Big/\sqrt{Z_\pi^\parallel(0)Z_\rho^\parallel(0)}=\frac{1}{e}\kappa_\text{exp}\approx 0.73337\;\text{GeV}^{-1}
\end{eqnarray}
where $e=\sqrt{4\pi\alpha}$. $Z_\pi^\parallel(0)$ and $Z_\rho^\parallel(0)$ are extracted from the definition \eqref{piparameter} and \eqref{rhoparameter}, while the quark propagator take the form without magnetic field. The tree level parameter is obtained according to $g^\text{tree}_{\rho\pi}=\frac{\kappa_\text{exp}}{e}\sqrt{Z_\pi^\parallel(0)Z_\rho^\parallel(0)}-g^\text{loop,vac}_{\rho\pi}=0.00068\;\text{GeV}^{-1}$. This shows that, within the local normalization of the bosonized scheme, the explicit loop contribution already accounts for most of the vacuum $\rho\pi\gamma$ coupling, leaving only a very small local remainder to be absorbed into the tree term.

\paragraph{Numerical result}  In the numerical calculation, we use the Pauli Villars regulation with three regulators $a_i=\{0,1,2,3\}$, $c_i=\{1,-3,3,-1\}$. The model parameters are chosen to be $\Lambda=1.25$GeV, $G_S=3.1$GeV$^{-2}$, $G_V=9.31974$GeV$^{-2}$ and $m_0 = 5$ MeV to reproduce the vacuum pion mass $0.135$GeV and $0.77$ GeV. One should notice that in the NJL RPA method the vector coupling $G_V=5.019$GeV$^{-2}$ is usually smaller in the same Pauli Villars regulation. In this derivative expansion method, however, the model requires a much larger coupling in the vector channel to reproduce the same rho mass in vacuum. This is because in the vacuum rho mass definition $\sqrt{(\frac{1}{2G_V}-\Pi_{\rho\rho}^{33})/Z_\rho^\parallel(0)}$, the loop contribution $\Pi_{\rho\rho}^{33}$ is evaluated at zero external momentum instead of the pole and this is quite the nature of the local expansion. In vacuum, the vector correlator has the transverse form $\Pi^{\mu\nu}(q)=(q^\mu q^\nu-q^2 g^{\mu\nu})\Pi(q^2)$, so $\Pi^{\mu\nu}(q)\propto q^2$, in particular $\Pi^{\mu\nu}(0)=0$. That means: a strictly local, zero-derivative loop term $\sim \rho^\mu \rho_\mu$ generated by a conserved vector current is constrained to be absent. In Pauli Villars regularization, those cancellations are often especially exact, so seeing the mass-type loop integral go to zero is actually consistent with the symmetry expectation. This implies the rho mass in a local bosonized EFT should come primarily from the tree piece, while the loop mainly builds the kinetic structure. So retuning $G_V$ to force $m_\rho^{pol}=0.77$GeV is usually a sign that the “bosonized local quadratic form” we are using for the rho is missing the correct momentum structure, not that QCD really wants $G_V$ that large. 

\begin{figure}[H]
\centering
\includegraphics[width=0.4\textwidth]{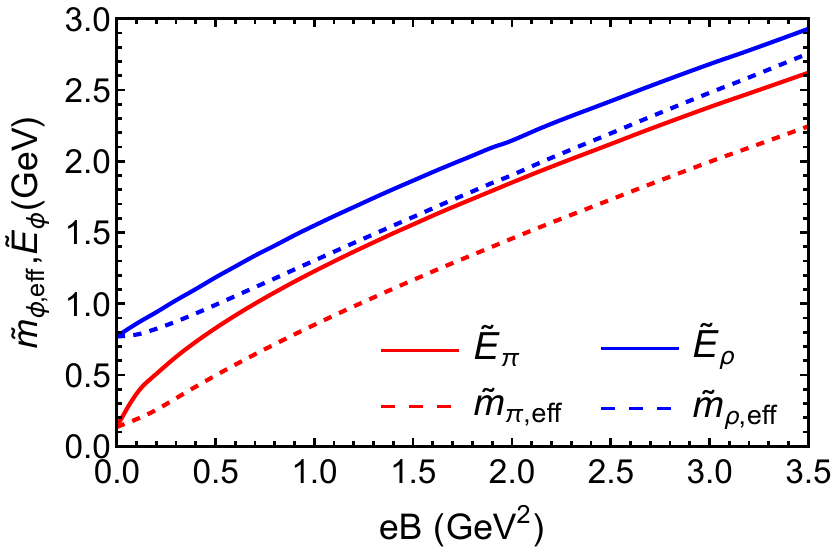}
\caption{
Local bosonized extraction without mixing. The solid lines show the unmixed LLL energies of the pion and longitudinal rho modes obtained from the local derivative-expanded quadratic action $\tilde{E}_{\phi }^2=(Z_\phi^\perp |eB|+m_{\phi,\text{eff}}^2)/ Z_{\phi}^\parallel$. The dashed lines show the corresponding effective mass parameters, $\widetilde m_{\phi,{\rm eff}}^2=m_{\phi,{\rm eff}}^2/Z_\phi^\parallel$. In this local scheme, the unmixed charged modes remain monotonic functions of the magnetic field.}
\label{boson_Em}
\end{figure}

For the chosen parameter set, the derivative expansion yields magnetic field dependent kinetic coefficients and effective mass parameters. We first present the unmixed pion and longitudinal rho meson mass defined by $\tilde{m}_{\pi,\text{eff}}^2=m_{\pi,\text{eff}}^2/ Z_{\pi }^\parallel$ and $\tilde{m}_{\rho,\text{eff}}^2=m_{\rho,\text{eff}}^2/ Z_{\rho }^\parallel$, together with the corresponding LLL energies. Similar to results in Secs. \ref{sec_rest} and \ref{sec_determinant} the unmixed LLL energies all exhibit monotonic dependence on magnetic field. 

\begin{figure}[H]
\centering
\includegraphics[width=0.4\textwidth]{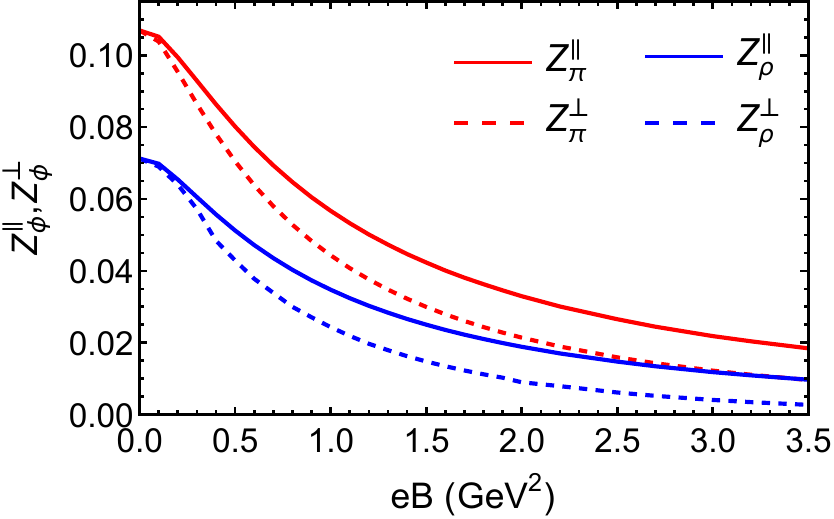}
\includegraphics[width=0.4\textwidth]{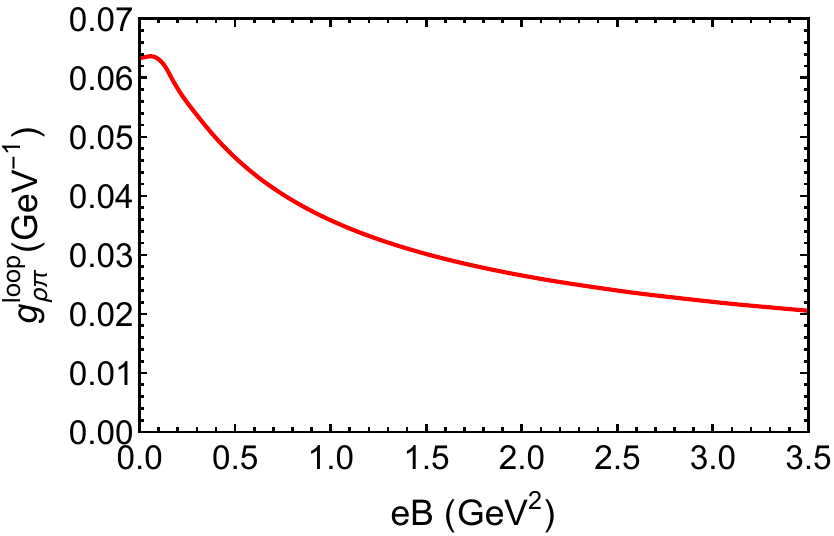}
\caption{
Local bosonized coefficients as functions of the magnetic field. Left panel: longitudinal and transverse wave-function renormalization factors $Z_\phi^\parallel$ and $Z_\phi^\perp$ for the charged pion and longitudinal charged rho mode. Right panel: loop-induced mixing coefficient $g_{\rho\pi}^{\rm loop}=H_{\rho\pi}/eB$ extracted from the local derivative expansion. The suppression of the kinetic coefficients enhances the normalized mixing, while the loop coefficient itself decreases with increasing magnetic field.}
\label{boson_gZ}
\end{figure}

In the left panel we present the anisotropic wave-function renormalization of charged pion and longitudinal rho meson. The magnetic field breaks rotational symmetry and separates longitudinal and transverse dynamics. $Z^\perp$ and $Z^\parallel$ are identical at vanishing magnetic field, due to the restoration of Lorentz symmetry. As $eB$ increases, the transverse quark wave functions become increasingly localized in the plane perpendicular to the field, and the polarization integrals become dominated by the lowest Landau levels. This reduces the sensitivity of the polarization function to the external frequency, leading to the strong suppression of both $Z_\pi$ and $Z_\rho$. It is worth noticing that in the local extraction, the wave-function renormalization is also extracted at $q=0$ instead of the pole.

The loop-order mixing coupling $g^{\text{loop}}_{\rho\pi}$ defined in \eqref{parameters} is presented as a function of $eB$ in the left panel of Fig.\ref{boson_gZ}. In contrast to the constant tree-level contribution, the loop-induced mixing originates from Landau-level sums involving Laguerre polynomials and phase-space integrals. The increasing localization of transverse wave functions and the reduction of available phase space lead to partial cancellations among Landau level contributions. As a result, the loop mixing coupling decreases with increasing magnetic field. 

\begin{figure}[H]
\centering
\includegraphics[width=0.4\textwidth]{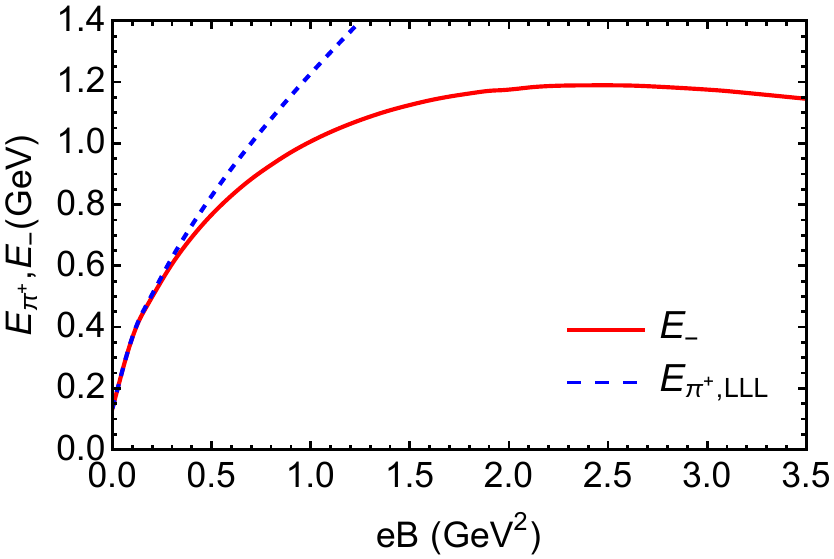}
\caption{
Lower mixed eigenmode in the local bosonized scheme. The red solid line shows the lower eigenvalue $E_-$ of the coupled $\pi^+-\rho_L^+$ system obtained from the local derivative-expanded quadratic action. The blue dashed line shows the unmixed charged-pion LLL energy. The local scheme produces a turnover only at relatively large magnetic field.}
\label{boson_eigen}
\end{figure}

We then compare the lower eigenmode $E_-$ of the mixed $\pi^+-\rho_L^+$ system obtained from \eqref{bosonization_eigen} with the unmixed charged-pion energy in the lowest Landau level. The result is shown in Fig. \ref{boson_eigen}. At small magnetic field, the mixed eigenvalue remains close to the unmixed pion mode. As $eB$ increases, the lower branch eventually develops a non-monotonic behavior. Compared with the pole-based schemes discussed later, however, the effect is much weaker here and sets in only at relatively large magnetic field.

This weaker behavior can be understood from the structure of the effective mixing in the local scheme. The relevant quantity entering the canonically normalized quadratic Lagrangian is $h(B)\sim (g^{\text{loop}}_{\rho\pi}+g^{\text{tree}}_{\rho\pi})eB(Z_\pi^\parallel Z_\rho^\parallel)^{-1/2}$. Although the local wave-function renormalization factors 
$Z_\pi^\parallel(B)$ and $Z_\rho^\parallel(B)$  decrease with increasing magnetic field and therefore enhance the effective mixing, the tree-level contribution in this scheme is numerically very small, while the loop-induced mixing itself does not grow with B. As a result, the residue enhancement is not sufficient to generate a strong turnover at moderate field, and the downward bending of the lower mode appears only much later.

The local bosonized scheme therefore still captures part of the underlying mechanism: residue suppression can amplify $\pi-\rho$ mixing and eventually distort the lowest mode. At the same time, the present result shows that this effect is not robust within a strictly local derivative expansion. In this approach, the turnover is delayed and quantitatively weak, indicating that the local expansion does not fully retain the charged pole structure relevant for the lattice observable. This motivates the next step of the analysis: returning to the nonlocal quadratic kernel itself and extracting the spectrum by imposing Landau-level projection directly on the external meson state.

\section{Direct determinant with Landau projection}
\label{sec_determinant}
We now return to the full nonlocal quadratic kernel and extract the charged excitation energy by solving it directly in a Landau-level basis. This step goes beyond both the rest-mass and local-expansion schemes. Unlike the rest-mass construction, the external meson is no longer treated as a plane-wave mode in the rest frame; unlike the local bosonized approach, no derivative expansion around small momentum is introduced. Instead, the relevant charged mode is projected directly onto a fixed Landau level before the pole condition is imposed.

For a charged particle in a magnetic field, the physical transverse eigenstates are Landau levels rather than plane waves. The quantity most closely related to the lattice observable is therefore not a rest mass, but the energy of the lowest Landau eigenmode of the coupled system. This motivates solving
\begin{eqnarray}
\label{det_LLL}
\text{det}\mathcal{K}_{LLL}(q_0)=0,
\end{eqnarray}
with the quadratic kernel projected onto the lowest Landau level of the charged meson sector.

This construction answers a more directly physical question: what is the energy of the lowest charged Landau eigenmode in the coupled $\pi-\rho$ system? Since the pole is obtained from the full Landau-projected kernel, the method retains the nonlocal structure of the microscopic dynamics while incorporating the correct kinematics of the external charged meson state.

For a charged relativistic scalar mode in a constant magnetic field, the dispersion relation takes the form $E_n^2=m^2+p_z^2+(2n+1)eB$. For mesons at rest along the field direction $p_z=0$, the physical energy of the lowest Landau mode is $E_{\text{LLL}}=\sqrt{m^2+eB}$. This quantity is directly related to the energy extracted from long-time decay of charged correlators in lattice simulations. To implement this kinematics consistently in the coupled $\pi-\rho$ system, the quadratic kernel must be evaluated between Landau-level wavefunctions of the external meson states. To derive the quadratic action in Landau-level basis for fixed Landau level n with $q_z$=0, we expand the charged meson fields in Landau eigenmodes:
\begin{eqnarray}
\pi(x)&=\sum_{n,q_y,q_z}\pi_n(q_0,q_3)\varphi_{n,q_1}(x_\perp)e^{-iq_0t+iq_3z},\nonumber\\ 
\rho_3(x)&=\sum_{n,q_y,q_z}\rho_n(q_0,q_3)\varphi_{n,q_1}(x_\perp)e^{-iq_0t+iq_3z}. 
\end{eqnarray}
Here $\varphi_{n,q_1}$ are the standard gauge dependent Landau wavefunctions. In the above expression we keep $q_3$ for completeness. In the calculation, one should restrict to $q_3=0$ to avoid coupling between $\rho_3$ and $\rho_0$. Then the boson quadratic action becomes a sum over Landau levels:
\begin{eqnarray}
\label{projected_kernel}
S^{(2)}_\text{eff}
&=&\frac{1}{2}\sum_n\int\frac{dq_0}{2\pi}
\Phi_n^\dagger(q_0)
\mathcal K(q_0;B,n)
\Phi_n(q_0),\nonumber\\
\mathcal{K}(q_0;B,n)&=&\left(\begin{array}{cc}
\mathcal{K}_{\pi\pi}(q_0;B,n)  &  \mathcal{K}_{\pi\rho}(q_0,B,n) \\
 \mathcal{K}_{\rho\pi}(q_0;B,n)  &    \mathcal{K}_{\rho\rho}(q_0;B,n) \\
\end{array}\right),
\end{eqnarray}
with $\Phi_n=(\pi_n,\rho_n)^T$, and $\mathcal{K}(q_0;B,n)$ is the $2\times 2$ Landau-projected kernel of Landau level n. The different components of the kernel are connected to the polarization function by 
\begin{eqnarray}
\label{kernel}
\mathcal{K}_{\pi\pi}&=&\frac{1}{2G_S}-\Pi_{\pi\pi,n}(q_0,B),\nonumber\\
\mathcal{K}_{\rho\rho}&=&\frac{1}{2G_V}-\Pi_{\rho\rho,n}(q_0,B),\nonumber\\
\mathcal{K}_{\rho\pi}&=&-{\Pi}^3_{\rho\pi,n}(q_0,B)+\mathcal{K}_{\rho\pi}^{\text{tree}}(q_0,B),\nonumber\\
\mathcal{K}_{\pi\rho}&=&-{\Pi}^3_{\pi\rho,n}(q_0,B)+\mathcal{K}_{\pi\rho}^{\text{tree}}(q_0,B).
\end{eqnarray}
One should notice that in the kernel $\Pi_{\pi\pi,n}(q_0,B)$ and $\Pi_{\rho\rho,n}(q_0,B)$ are different from the ordinary RPA polarization function in last section. In the rest-mass scheme, the pole condition determines the mass of the composite bound state. It is the energy cost to create a pion at rest. It does not describe the motion of a charged pion in an external magnetic field. On the other hand the kernel $\mathcal{K}_{\pi\pi}(q_0;B,n)$ is evaluated in a Landau-level basis for the pion field. $\Pi_{\pi\pi,n}(q_0,B)$ is the quark loop projected onto the pion Landau mode n, i.e. with the appropriate Landau wavefunctions on the external pion legs. This “projection onto n” is exactly what is missing when one computes $\Pi(q)$ as a function of continuous $q_\perp$ and then sets $q_\perp=0$. The key point is that Landau levels enter through the external pion mode projection, not through the internal quark Landau sums alone. The “Landau-projected” polarization function is
\begin{eqnarray}
\Pi_{\phi_1\phi_2,n}(q_0,B)=\int d^4xd^4y \varphi_{n}^*(x)\Pi_{\phi_1\phi_2}(x,y;B)\varphi_{n}(y),
\end{eqnarray}
where $\Pi_{\phi_1\phi_2}(x,y;B)$ is the coordinate-space quark loop, for charged pion and rho this quark loop has non-vanishing Schwinger phases. $\varphi_{n}(r)$ is the standard wave function of charged pion at Landau level n, which is given by
\begin{eqnarray}
\label{wavefunction}
\varphi_{n}(r)=e^{-ip_0 t+ip_3z}\varphi_{n,p_1}(\vec{r}_\perp)=e^{-ip_0 t+ip_3z}\frac{{\sqrt{2\pi}}}{\sqrt{2\pi l}}\frac{e^{-\frac{\left(\frac{y}{l}+s_\perp p_1 l\right)^2}{2}}}{\sqrt{2^nn!\sqrt{\pi}}}H_n\left(\frac{y}{l}+s_\perp p_1 l\right)e^{ip_1x},
\end{eqnarray}
where $l=1/\sqrt{eB}$ is the magnetic length. $H_n(x)$ is the Hermite polynomials, and $s_{\perp}\equiv \text{sign}(eB)$. The wave functions satisfy the conditions of normalizability and completeness. The “Landau-projected” polarization functions in the kernel \eqref{kernel} are defined through
\begin{eqnarray}
\label{chargedloop}
\Pi_{\pi\pi,n}&=&-i\int_{r,r'}\text{Tr}\Big[\varphi^*_{n}(r')(i\gamma^5\tau^-)S_f(r,r')(i\gamma^5\tau^+)S_f(r',r)\varphi_{n}(r)\Big],\nonumber\\
\Pi_{\rho\rho,n}&=&-i\int_{r,r'}\text{Tr}\Big[\varphi^*_{n}(r')(\gamma^3\tau^-)S_f(r,r')(\gamma^3\tau^+)S_f(r',r)\varphi_{n}(r)\Big],\nonumber\\
\Pi_{\rho\pi,n}^3&=&-i\int_{r,r'}\text{Tr}\Big[\varphi^*_{n}(r')(\gamma^3\tau^-)S_f(r,r')(i\gamma^5\tau^+)S_f(r',r)\varphi_{n}(r)\Big],\nonumber\\
\Pi_{\pi\rho,n}^3&=&-i\int_{r,r'}\text{Tr}\Big[\varphi^*_{n}(r')(i\gamma^5\tau^-)S_f(r,r')(\gamma^3\tau^+)S_f(r',r)\varphi_{n}(r)\Big].
\end{eqnarray}
We focus on the case where the external pion and rho are in the lowest Landau level, namely take $n=0$ in the above expressions. After taking all the integrals, we finally arrive at 
\begin{eqnarray}
\label{polarization}
\Pi_{\pi\pi,0}&=&-\frac{4N_c eB}{3\pi}\int\frac{dk_3}{(2\pi)}\sum_{n,m=0}^\infty \Big[Y_1(n,m) I_3(q_0)+[(k_3^2+m^2) Y_1(n,m)+\frac{8}{3l^2}Y_2(n,m)]I_2(q_0)\Big],\nonumber\\
\Pi_{\rho\rho,0}&=&-\frac{4N_c eB}{3\pi}\int\frac{dk_3}{(2\pi)}\sum_{n,m=0}^\infty \Big[Y_1(n,m) I_3(q_0)+[(-k_3^2+m^2) Y_1(n,m)+\frac{8}{3l^2}Y_2(n,m)]I_2(q_0)\Big],\nonumber\\
\Pi_{\rho\pi,0}^3&=&-(-im_fq_0)\frac{4N_c eB}{3\pi}\int\frac{dk_3}{(2\pi)}\sum_{n,m=0}^\infty Y_3(n,m) I_2(q_0)~\equiv~-iq_0\;eB\;g_{\rho\pi}^\text{loop},\nonumber\\
\Pi_{\pi\rho,0}^3&=&-(+im_fq_0)\frac{4N_c eB}{3\pi}\int\frac{dk_3}{(2\pi)}\sum_{n,m=0}^\infty Y_3(n,m) I_2(q_0)
~\equiv~+iq_0\;eB\;g_{\rho\pi}^\text{loop},
\end{eqnarray}
with 
\begin{eqnarray}
\label{gloop}
g_{\rho\pi}^\text{loop}&\equiv&-\frac{4N_c}{3\pi}m_f\int\frac{dk_3}{(2\pi)}\sum_{i,j=0}^\infty Y_3(n,m) I_2(q_0).
\end{eqnarray}
Obviously, hermiticity gives $\mathcal{K}_{\rho\pi}=\mathcal{K}_{\pi\rho}^*$. The functions $I_2(q_0)$ and $I_3(q_0)$ encodes the Matsubara sum in the quark loop, the result are presented in appendix.\ref{App_matsubara}. The “Landau-projected” polarization function defined in \eqref{chargedloop} has complicated coordinate and momentum integrals, such integrals generates coefficiens $Y_i(n,m)$ in \eqref{polarization}, we include the details of calculation in appendix.\ref{App_project}. Same as in the rest-mass section,  the off-diagonal component of the kernel can be rewritten as $-iq_0H(q_0, B)\equiv -{\Pi}^3_{\pi\rho}+\mathcal{K}_{\pi\rho}^{\text{tree}}$ and $iq_0H(q_0, B)\equiv -{\Pi}^3_{\rho\pi}+\mathcal{K}_{\rho\pi}^{\text{tree}}$, with $H(B)=(g_{\rho\pi}^\text{loop}+g^\text{tree}_{\rho\pi}) eB$ in the kernel denotes the combined loop-induced and tree-level mixing coefficient, with $g^\text{tree}_{\rho\pi}$ determined in Sec.\ref{com_input}.

The spectrum is obtained by solving \eqref{det_LLL}, yielding two mixed eigenmodes $q_0=E_\pm(B)$. In this formulation, the Landau-level structure is incorporated directly in the polarization functions. The determinant has the generic structure $\mathcal{K}_{\pi\pi}(q_0,B)\mathcal{K}_{\rho\rho}(q_0,B)-q_0^2H^2(B)=0$. As the magnetic field increases, both the diagonal Landau-level energies and the off-diagonal mixing term grow with $eB$. Their competition produces level repulsion between the pion-like and rho-like modes.

\begin{figure}[H]
\centering
\includegraphics[width=0.4\textwidth]{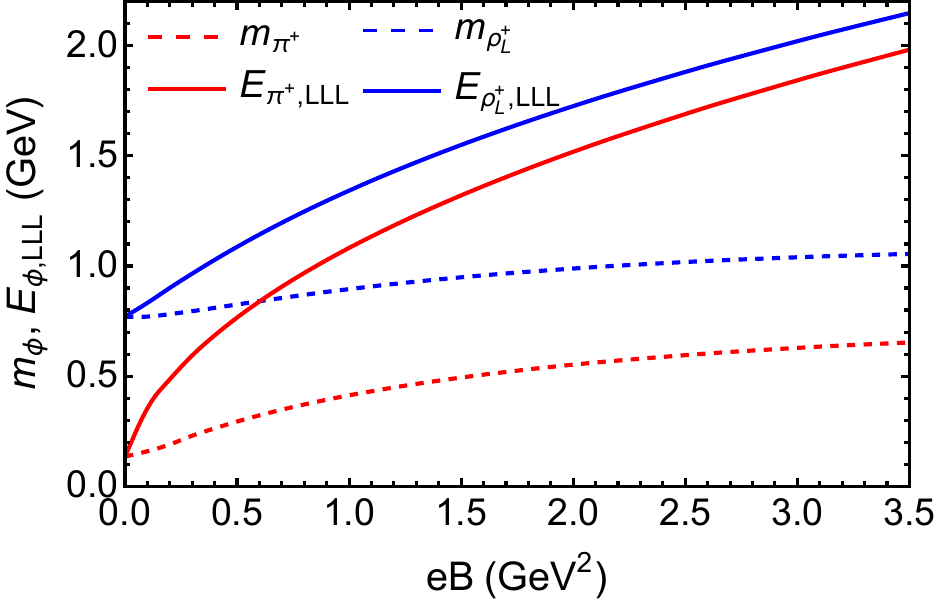}~~
\includegraphics[width=0.4\textwidth]{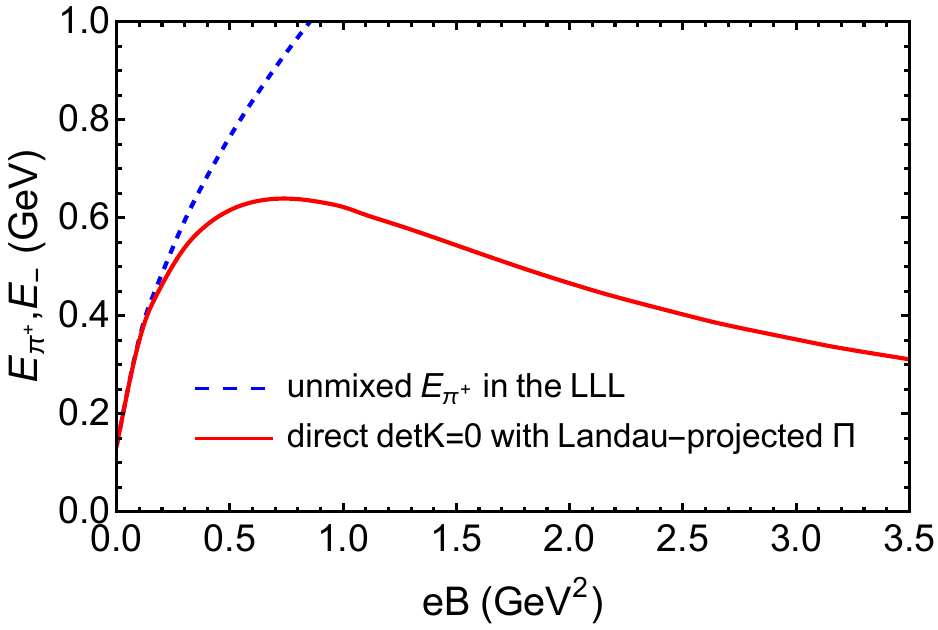}
\caption{
Direct determinant extraction with Landau projection. Left panel: unmixed pion and longitudinal rho energies obtained from the diagonal components of the Landau-projected kernel, together with the corresponding rest-mass parameters $m_\phi=(E_\phi^2-eB)^{1/2}$. Right panel: lower mixed eigenmode obtained by solving the full Landau-projected determinant, compared with the unmixed pion LLL energy. The direct determinant scheme yields a robust non-monotonic lowest mode.}
\label{detsolve1}
\end{figure}

Using the same parameter set presented in Sec.\ref{com_input}, we solve the determinant equation in the LLL sector.  Before introducing mixing, it is useful to examine the unmixed LLL energies extracted from the diagonal Landau-projected kernels, since these provide the direct baseline for the charged quasiparticle spectrum. The unmixed LLL energies of the charged pion and the longitudinally polarized charged rho meson are obtained from the diagonal Landau-projected kernels. As presented in the left panel in Fig.\ref{detsolve1}, the resulting pole energies $E_\pi(B)$ and $E_\rho(B)$ increase monotonically with the magnetic field and exhibit the expected $\sqrt{eB}$ behavior at large field strength, reflecting Landau quantization. When converted to rest masses via $m_\phi=(E_\phi^2-eB)^{1/2}$, the charged pion mass increases monotonically with $eB$, while the longitudinal rho mass shows a mild nonmonotonicity consistent with previous model studies \cite{Liu:2014uwa}. In the absence of mixing, therefore, the charged pion sector does not exhibit any turnover, indicating that the quark-antiquark binding and Landau quantization are insufficient to explain the lattice observations.

The resulting lowest eigenvalue from solving the pole condition \eqref{det_LLL} is presented in the right panel of Fig.\ref{detsolve1}. At intermediate magnetic fields, the repulsive mixing effect dominates over the diagonal growth of the pion channel, driving the lower eigenmode downward and generating a non-monotonic behavior. Unlike the rest-mass scheme, here the lowering of the eigenmode does not require a negative squared mass parameter; it emerges directly from a fully dynamical interplay between Landau quantization and $\pi-\rho$ mixing.

The direct-determinant scheme therefore provides the first extraction in which the quantity of interest is directly the energy of the lowest charged Landau eigenmode. Unlike the rest-mass construction, it does not rely on a reconstruction of Landau kinematics, and unlike the local-expansion scheme, it does not pass through a derivative-expanded mesonic EFT. The pole is obtained directly from the Landau-projected quadratic kernel, making this the most kinematically faithful scheme in the present comparison.

\section{Near-pole extraction (quasiparticle effective theory)}
\label{sec_nearpole}

While the direct-determinant scheme captures the charged Landau level kinematics most faithfully, it does not by itself make transparent why the mixing becomes so effective. To expose the mechanism more clearly, we keep the same Landau projected quadratic kernel but expand it near the physical poles and rewrite the system in terms of canonically normalized quasiparticle fields. This leads to the near-pole extraction, which provides the clearest interpretation of the turnover in terms of residue enhanced $\pi-\rho$ mixing.

In the near-pole scheme, the energy eigenvalues are extracted by expanding the Landau projected quadratic kernel around the unmixed physical poles and then canonically normalizing the corresponding modes. In this sense, the near-pole construction is built on the same Landau-level kinematics as the direct-determinant approach, but reorganizes the result into a quasiparticle effective theory. This makes it possible to identify directly how wave-function renormalization modifies the effective mixing strength.

For each of pion and longitudinal-rho channel, we want the Landau-projected quadratic kernel \eqref{projected_kernel} written near the physical pole as 
\begin{eqnarray}
K_{\phi\phi}(q_0; B, n)\approx Z(q_0^2-E^2(B,n)), 
\end{eqnarray}
where $E(B,n)$ is the unmixed pole energy, and $Z$ is the corresponding residue factor. In the present work we focus on the lowest Landau level, $n=0$, and restrict to $q_z=0$. First turn off mixing, and define energies by
\begin{eqnarray}
\label{pole}
&&\mathcal{K}_{\pi\pi}(q_0=E_\pi;B,n=0)=0,\qquad\qquad \mathcal{K}_{\rho\rho}(q_0=E_\rho;B,n=0) =0,
\end{eqnarray}
with the polarization function defined in \eqref{polarization} in Sec.\ref{sec_determinant}. These are the unmixed pion and rho energies at lowest Landau level. Crucially, near the pole, each diagonal kernel can be expanded in $q_0^2$, 
\begin{eqnarray}
\mathcal{K}_{\pi\pi,0}(q_0)\approx \frac{\partial \mathcal{K}_{\pi\pi,0}}{\partial (q_0^2)}\Big|_{q_0^2=E_\pi^2}(q_0^2-E_\pi^2),\qquad\qquad 
\mathcal{K}_{\rho\rho,0}(q_0)\approx \frac{\partial \mathcal{K}_{\rho\rho,0}}{\partial (q_0^2)}\Big|_{q_0^2=E_\rho^2}(q_0^2-E_\rho^2),
\end{eqnarray}
with the residues defining the wavefunction renormalizations,
\begin{eqnarray}
Z_\pi(B,n)\equiv \frac{\partial \mathcal{K}_{\pi\pi,0}}{\partial (q_0^2)}\Big|_{q_0^2=E_\pi^2},\qquad\qquad 
Z_\rho(B,n)\equiv \frac{\partial \mathcal{K}_{\rho\rho,0}}{\partial (q_0^2)}\Big|_{q_0^2=E_\rho^2}.
\end{eqnarray}
For the non-diagonal term in the kernel \eqref{projected_kernel}, we recall the loop-induced mixing $g_{\rho\pi}^\text{loop}$ defined in \eqref{gloop}, and the tree-level mixing $g^\text{tree}_{\rho\pi}$ is determined in Sec.\ref{com_input}. Together, they give the off-diagonal term $\mathcal K_{\rho\pi}(q_0,B)= +i q_0 H(B)$ and $\mathcal K_{\pi\rho}=\mathcal K_{\rho\pi}^*$, where $H(B)=(eB)(g^\text{loop}_{\rho\pi}+g^\text{tree}_{\rho\pi})$.

The canonically nomalized fields are then defined through $\pi_c\equiv \sqrt{Z_\pi}\pi$ and $\rho_c\equiv \sqrt{Z_\rho}\rho$. In the canonical basis $\Phi_c=(\pi_c,\rho_c)^T$, the near-pole LLL quadratic kernel is 
\begin{eqnarray}
\label{determinantK}
\mathcal{K}_c^{\text{LLL}}(q_0)\approx\left(\begin{array}{cc}
q_0^2-E_\pi^2  &  -i q_0 h(B) \\
i q_0 h(B) &    q_0^2-E_\rho^2 \\
\end{array}\right).
\end{eqnarray}
After canonical normalization at the pole, the quadratic kernel takes the form of a two-level system with an effective mixing strength
\begin{eqnarray}
\label{hB}
h(B)= \frac{H(B)}{\sqrt{Z_\pi Z_\rho}}=\frac{eB(g^\text{loop}_{\rho\pi}+g^\text{tree})}{\sqrt{Z_\pi Z_\rho}}.
\end{eqnarray}
This construction answers the following question: once the charged excitation is projected onto the physical Landau level, how does canonical normalization reshape the effective $\pi-\rho$ mixing and the resulting quasiparticle spectrum? In particular, the near-pole scheme makes explicit that the off-diagonal mixing is enhanced by the factor $(Z_\pi Z_\rho)^{-1/2}$, so that a strong suppression of the mesonic residues can substantially increase the level repulsion even when the underlying microscopic mixing kernel itself remains moderate.

The eigenmodes are determined by solving
\begin{eqnarray}
\label{polecondition}
\text{det}\; \mathcal{K}_c^{\text{LLL}}(q_0)=0.
\end{eqnarray}
The near-pole scheme differs conceptually from both the determinant-based approaches: It does not solve the full nonlocal kernel without normalization. 
Besides, compared with the local bosonization scheme, the near-pole extraction evaluates the residues at the physical pole rather than at vanishing momentum. This makes the enhancement directly tied to the physical quasiparticle normalization rather than to a derivative expansion, providing a clearer dynamical interpretation of the turnover. Because the effective mixing strength scales inversely with $(Z_\pi Z_\rho)^{-1/2}$, suppression of the residue can enhance level repulsion in this framework. Thus, this scheme makes explicit how wave-function renormalization modifies the strength of mixing in the coupled system.

The unmixed LLL energies $E_\pi(B)$ and $E_\rho(B)$ are the same Landau-projected pole energies already obtained in Fig.\ref{detsolve1} of Sec.\ref{sec_determinant} from the diagonal kernel. The resulting pole energies $E_\pi(B)$ and $E_\rho(B)$ increase monotonically with the magnetic field and exhibit the expected $\sqrt{eB}$ behavior at large field strength. They are repeated here only to emphasize that the near-pole scheme does not introduce different kinematics, but rather a different organization of the same pole structure.

\begin{figure}[H]
\centering
\includegraphics[width=0.4\textwidth]{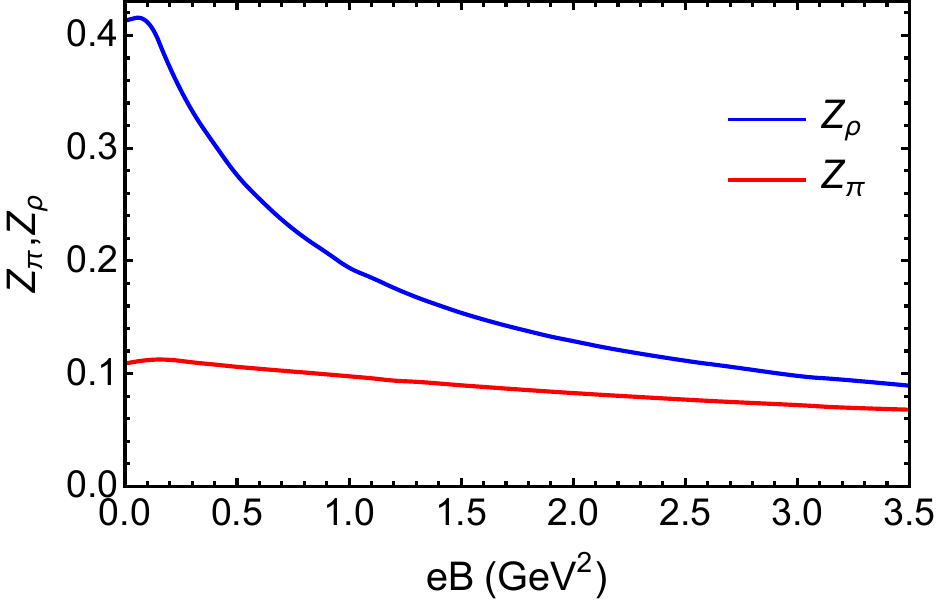}
\caption{
Near-pole wave-function renormalization factors extracted from the slopes of the Landau-projected diagonal kernels at the unmixed physical poles. The pion residue changes moderately, whereas the longitudinal rho residue is strongly suppressed with increasing magnetic field. This suppression is a key source of the enhanced effective mixing in the near-pole formulation.}
\label{nearpole_Z}
\end{figure}

The wave-function renormalizations extracted from the near-pole slopes of the diagonal kernels are presented in Fig.\ref{nearpole_Z}. The crucial property of the wave-function renormalizations is the qualitatively different magnetic-field dependence for the pion and rho channels. While $Z_\pi(B)$ (red line) decreases slowly with increasing field, $Z_\rho(B)$ (blue line) is rapidly suppressed in the LLL and becomes very small at strong magnetic fields. This behavior originates from the tensor structure of the longitudinal vector correlator in a magnetic background, leading to a steep reduction of the near-pole residue. This collapse of $Z_\rho(B)$ plays a central dynamical role in amplifying mixing effects near the pole.

\begin{figure}[H]
\centering
\includegraphics[width=0.4\textwidth]{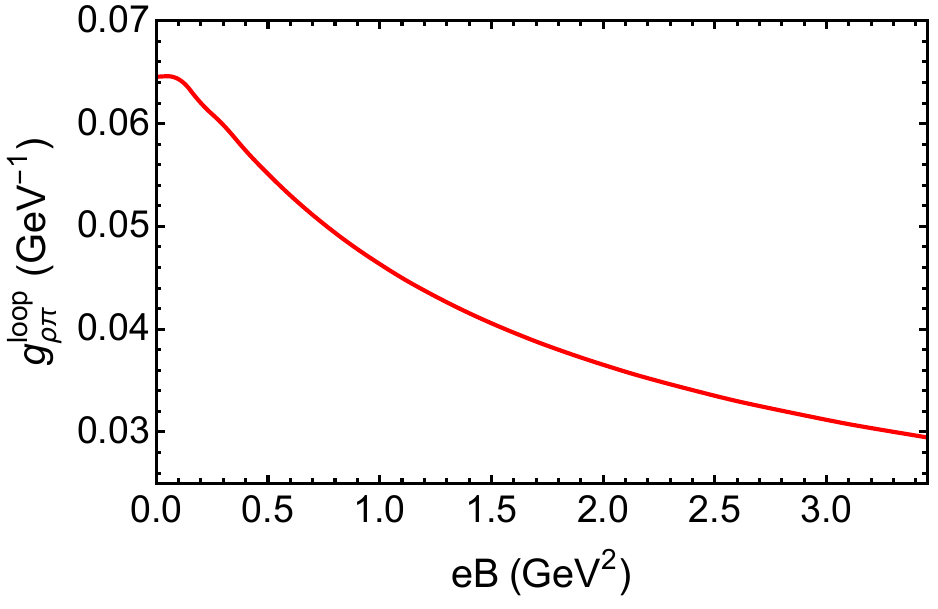}
\includegraphics[width=0.4\textwidth]{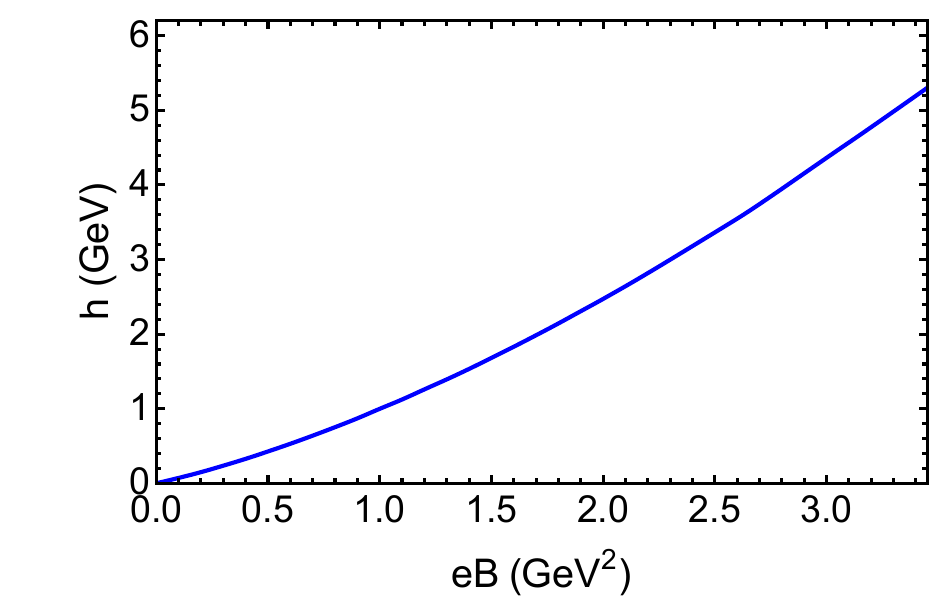}
\caption{
Mixing strength in the near-pole scheme. Left panel: loop-induced coefficient $g_{\rho\pi}^{\rm loop}$ extracted from the Landau-projected off-diagonal kernel and evaluated at the pion pole. Right panel: canonically normalized mixing strength $h(B)=eB(g_{\rho\pi}^{\rm loop}+g_{\rho\pi}^{\rm tree})/\sqrt{Z_\pi Z_\rho}$. The rapid growth of $h(B)$ is driven mainly by the suppression of the pole residues rather than by an increase of the microscopic loop coefficient.}
\label{nearpole_gandh}
\end{figure}

The loop-induced mixing coupling $g_{\rho\pi}^{\rm loop}(B)$, defined in \eqref{gloop}, is displayed in the left panel of Fig.\ref{nearpole_gandh}. Its B-dependence is rather moderate and decreases slowly with increasing magnetic field. This shows that the turnover in the near-pole scheme is not driven by a rapid growth of the microscopic loop-induced mixing itself. Instead, the relevant quantity is the canonically normalized mixing strength $h(B)$ \eqref{hB} shown in the right panel of Fig.\ref{nearpole_gandh}. Owing to the explicit $eB$ factor and, more importantly, the strong suppression of the pole residues, especially in the rho channel, $h(B)$ increases rapidly with $eB$. The near-pole scheme therefore makes transparent that the non-monotonic behavior is not controlled primarily by the size of the bare mixing kernel, but by residue-enhanced level repulsion in the canonically normalized quasiparticle basis.

\begin{figure}[H]
\centering
\includegraphics[width=0.4\textwidth]{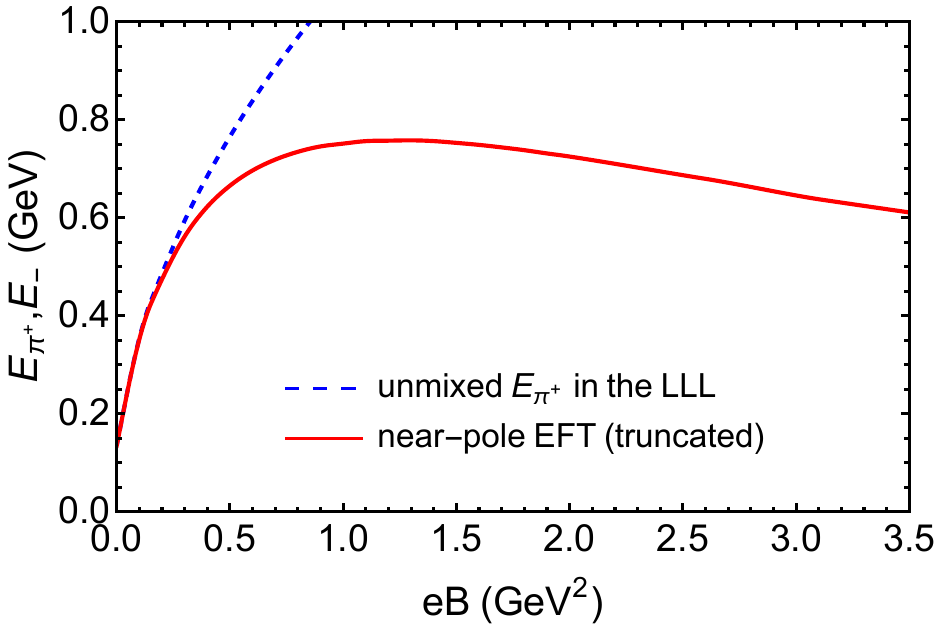}
\caption{
Lower mixed eigenmode in the near-pole quasiparticle scheme. The dashed blue line shows the unmixed pion LLL energy, while the red solid line shows the lower eigenmode obtained from the canonically normalized near-pole kernel. The turnover reflects residue-enhanced level repulsion between pion-like and rho-like Landau modes.}
\label{fig_nearpole}
\end{figure}

Solving the pole condition \eqref{polecondition} of the near-pole kernel yields two eigenmodes corresponding to mixed pion-rho states. The lower eigenmode $E_-$ is presented as the blue line in Fig.\ref{fig_nearpole}. At weak fields, the lower eigenmode closely tracks the unmixed pion LLL energy. At intermediate fields, increasing mixing induces strong level repulsion, causing the lower eigenvalue to bend downward. This behavior is a direct consequence of the residue enhanced mixing between a light pion and a heavier longitudinal rho mode in the same Landau level.

The near-pole scheme therefore provides the clearest dynamical interpretation of the non-monotonic behavior: the turnover is driven by residue enhanced level repulsion between pion like and rho like Landau modes. Its limitation is that the expansion assumes the mixed pole remains sufficiently close to the unmixed one. At very large magnetic field, where the pole shift becomes substantial, the direct determinant solution of the full Landau projected kernel is quantitatively more reliable. Taken together, the two schemes play complementary roles: the direct-determinant approach gives the most faithful charged mode spectrum, while the near-pole expansion makes the underlying mechanism most transparent.

\section{Comparative Analysis}
\label{sec_compare}
We now compare the four extraction schemes introduced above: rest mass reconstruction, local bosonization, direct determinant solving with Landau projection, and near-pole expansion. All four are derived from the same microscopic NJL kernel and differ only in how the same coupled $\pi-\rho$ dynamics are projected onto physical masses or energies. Their differences therefore do not reflect different underlying physics, but different ways of representing the same charged meson system in a magnetic background.

The comparison reveals a clear hierarchy. The rest mass and local expansion schemes provide useful reference descriptions, but neither gives a fully robust account of the lattice type turnover. By contrast, the direct-determinant and near-pole schemes, both built on the Landau projected charged pole structure, retain a pronounced non-monotonic lowest mode. Among them, the direct-determinant approach is the most kinematically faithful, while the near-pole scheme makes the mechanism of residue enhanced mixing most transparent.

\begin{figure}[H]
\centering
\includegraphics[width=0.4\textwidth]{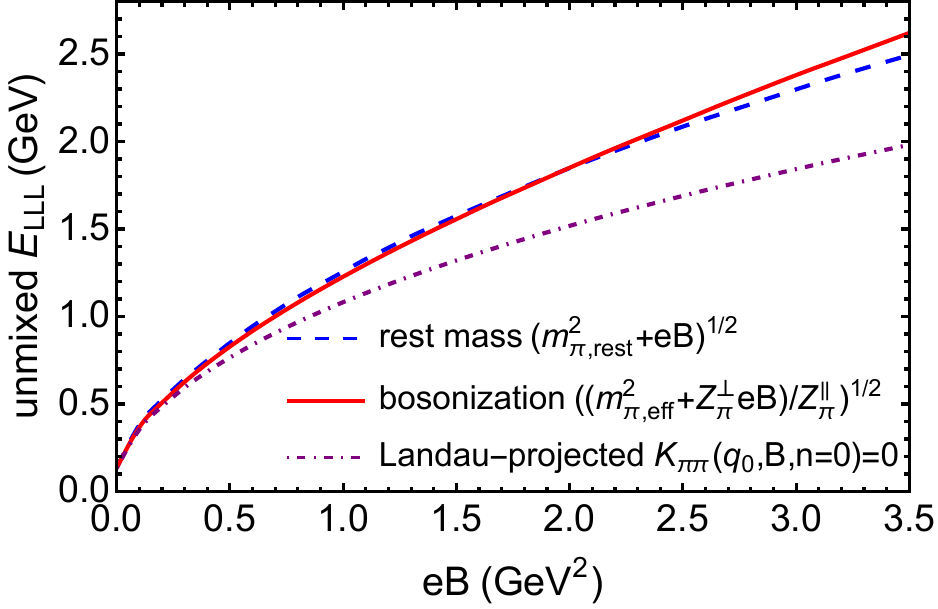}
\caption{
Unmixed charged-pion LLL energy obtained from different extraction schemes. The rest-mass and local bosonized schemes reconstruct the charged energy from vacuum-style or local effective quantities, while the Landau-projected determinant scheme extracts the charged pole directly from the LLL-projected microscopic kernel. The near-pole scheme uses the same unmixed Landau-projected energy as the direct determinant approach.}
\label{compare_unmixed}
\end{figure}

We first compare the unmixed charged pion energy in the lowest Landau level, as shown in Fig. \ref{compare_unmixed}. The rest mass and local bosonization schemes both start from vacuum style or small momentum constructions and only introduce the Landau level contribution at the level of an effective mesonic description. As a result, they lead to similar LLL energies. By contrast, in the direct determinant scheme with Landau-projected kernel the external pion is projected onto the Landau basis already at the level of the microscopic kernel, so the resulting LLL energy is lower and more directly tied to the charged quasiparticle pole. The near-pole scheme shares the same unmixed Landau projected energies, since it is built from the same diagonal kernel and differs only in how the pole is reorganized near the physical excitation.

\begin{figure}[H]
\centering
\includegraphics[width=0.4\textwidth]{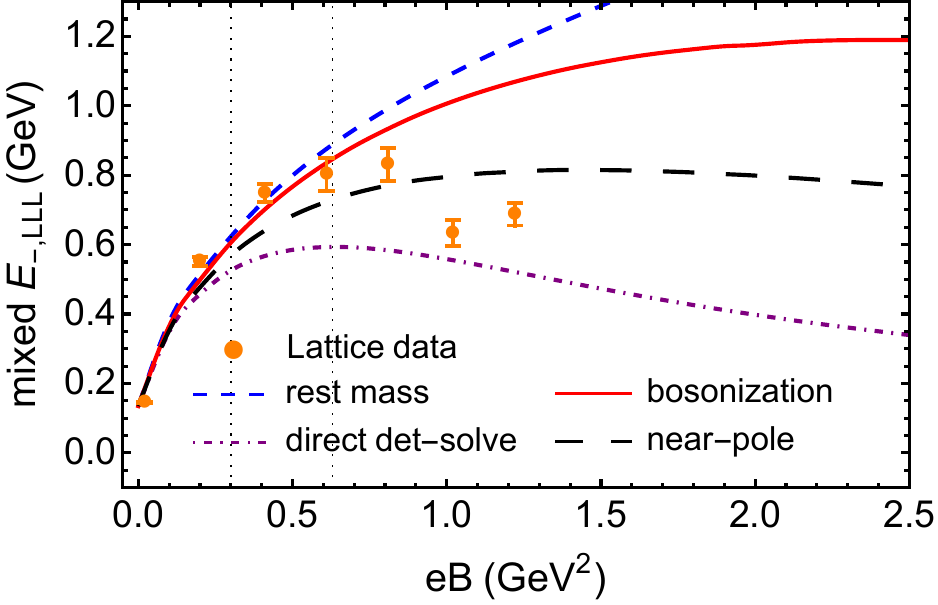}
\caption{
Comparison of the lower mixed eigenmode obtained from the four extraction schemes: rest-mass reconstruction, local bosonization, direct determinant solving with Landau projection, and near-pole expansion. The orange points with error bars denote the continuum-extrapolated charged-pion lattice data from Ref.~\cite{Ding:2026qzu}, shown as a reference for the lattice observable. The vertical dotted lines mark the field region where the lattice data start to deviate from the simple LLL trend and where the turnover occurs.}
\label{compare_mixed}
\end{figure}

The comparison of the mixed lower eigenmode is shown in Fig.~\ref{compare_mixed}, where the lattice data are also included as a direct reference. At weak magnetic field, all four schemes remain close to each other and to the lattice points, reflecting the fact that the mixing is still modest and the charged mode is dominated by the pion-like lowest Landau level. The difference among the schemes becomes evident in the intermediate field region, where the lattice data begin to deviate from the simple LLL trend and develop a turnover.

The rest-mass reconstruction fails to reproduce this behavior. Although the mixed rest mass itself can decrease with increasing magnetic field, the physical LLL energy reconstructed as $E_{\rm LLL}=(m_{\rm rest}^2+eB)^{1/2}$ remains too rigid and continues to rise. This makes the rest-mass scheme incompatible with the lattice trend once the data bend downward.

The local bosonized scheme generates a non-monotonic lower mode, but the turnover is delayed and the energy remains substantially above the lattice data in the intermediate-field region. This indicates that a local derivative expansion can retain part of the mixing mechanism, but does not fully capture the charged pole structure that controls the lattice observable.

By contrast, the two Landau-projected pole-based schemes show the closest qualitative connection to the lattice result. The direct determinant solution produces a turnover in approximately the same magnetic-field range as the lattice data, showing that the non-monotonic behavior can already emerge from the full Landau-projected quadratic kernel. The near-pole scheme also gives a robust decreasing tendency after the peak, and makes transparent that the downward bending is driven by residue-enhanced level repulsion. Quantitatively, the direct determinant curve give the most close turnover region compared to lattice result, while the near-pole curve tends to bend downward at larger fields because it relies on an expansion around the unmixed poles. Thus Fig.~\ref{compare_mixed} supports the interpretation that the lattice behavior is tied to the Landau-projected charged quasiparticle pole, rather than to a vacuum-style rest mass or to a strictly local mass parameter.

\paragraph{Rest-mass extraction}
In the rest-mass scheme, the pole is first determined at $\vec{q}=0$, and the LLL energy is reconstructed afterward through $E_{LLL}=\sqrt{m_\text{rest}^2+eB}$. As long as the extracted rest-mass squared remains positive, this immediately imposes a lower bound of order $\sqrt{eB}$ on the charged excitation energy. The scheme therefore cannot reproduce the lattice observation that the lowest mode bends downward at sufficiently large magnetic field. Its limitation is structural: the charged quasiparticle is not treated as a Landau eigenmode from the outset.

\paragraph{Local bosonization}
In the local bosonized scheme, the quadratic kernel is reorganized into a derivative expanded mesonic effective theory, and canonical normalization is implemented through local kinetic coefficients extracted at vanishing external momentum. This framework can still accommodate the basic mechanism that residue suppression enhances the effective $\pi-\rho$ mixing. However, in the present calculation the matched tree-level contribution is numerically very small in this scheme, and the resulting turnover becomes weak and delayed. The local expansion therefore captures part of the underlying physics, but does not provide a robust description of the lattice type non-monotonic behavior.

\paragraph{Direct determinant solving}
The direct determinant scheme solves the pole condition $\text{det}\,\mathcal{K}(q_0,B,n=0)=0$ in the Landau projected charged sector. No vacuum style dispersion relation is imposed, and no local derivative expansion is introduced. The resulting eigenvalue is therefore directly the energy of the lowest charged Landau mode. This makes the method the most kinematically faithful scheme in the present comparison. The resulting turnover appears in the same magnetic-field region as the lattice trend and gives the closest quantitative description among the four schemes around the peak.

\paragraph{Near-pole extraction}
The near-pole scheme starts from the same Landau projected kernel as the direct determinant approach, but expands it around the unmixed poles and rewrites the problem in terms of canonically normalized quasiparticle fields. Its main advantage is interpretive rather than kinematic: it makes explicit that the effective mixing scales as $h(B)\propto (g^\text{tree}_{\rho\pi}+g^\text{loop}_{\rho\pi})eB(Z_\pi Z_\rho)^{-1/2}$, so that the strong suppression of the pole residues enhances the level repulsion. In this way, the near-pole scheme provides the clearest dynamical explanation of why the lowest mode turns over, even though at very large magnetic field the direct determinant solution remains quantitatively more stable.

Taken together, the four schemes reveal a consistent physical picture. The non-monotonic behavior is not a generic consequence of introducing $\pi-\rho$ mixing into any effective description. Rather, it emerges robustly only in schemes that retain the charged Landau projected pole structure. The rest-mass construction misses the effect because its dispersion relation is too rigid, while the local bosonized expansion retains only a weakened version of the mechanism. The direct determinant solution captures the charged quasiparticle spectrum most faithfully, and the near-pole expansion shows most transparently that the turnover is driven by residue enhanced level repulsion between pion like and rho like modes. In this sense, the two Landau projected pole-based schemes provide the most reliable interpretation of the lattice result.

\section{Implications for Lattice and EFT}
\label{sec_implic}
The comparison of extraction schemes clarifies what is actually probed in lattice simulations of charged mesons in a magnetic field. Lattice calculations extract the lowest energy eigenmode from Euclidean correlation functions. For a charged excitation, this quantity is most naturally associated with the pole of the Landau projected propagator, rather than with a rest mass parameter defined at vanishing spatial momentum and then supplemented by a simple dispersion relation. In this sense, the direct determinant and near-pole schemes provide the closest theoretical counterparts to the lattice observable.

This comparison also shows that the non-monotonic behavior of the charged pion mode should not be interpreted as a generic consequence of introducing $\pi-\rho$ mixing into any effective description. The rest-mass construction fails to reproduce the turnover because its dispersion relation is too rigid, while the local bosonized expansion retains only a weakened and delayed version of the effect. By contrast, the turnover remains robust in the two Landau projected pole-based schemes. The lattice result should therefore be understood not simply as evidence for mixing, but more specifically as evidence for a quasiparticle mixing phenomenon in which the charged pole structure in a magnetic field is essential.

From the effective field theory point of view, the analysis highlights two requirements. First, the effective description must contain the gauge-invariant $\pi-\rho-$electromagnetic coupling that allows mixing in a magnetic background. Second, it must treat the normalization of the charged modes consistently. The present results show that the size of the microscopic mixing kernel alone is not enough to determine the spectrum. What matters physically is the canonically normalized mixing strength, which can be strongly enhanced when the mesonic residues are suppressed.

The near-pole analysis makes this point especially transparent. There, the effective mixing scales as $h(B)\propto (g^\text{tree}_{\rho\pi}+g^\text{loop}_{\rho\pi})eB(Z_\pi Z_\rho)^{-1/2}$, so that the turnover is driven not simply by the growth of the bare mixing term, but by residue enhanced level repulsion between pion like and rho like charged modes. At the same time, the direct determinant result shows that this mechanism is not merely a model dependent reinterpretation: it is already encoded in the full Landau projected quadratic kernel itself.

These observations also clarify the role of locality. A local bosonized expansion can still reproduce part of the mechanism, because the suppression of local kinetic coefficients enhances the effective mixing there as well. However, the present calculation shows that this effect is quantitatively weak and delayed in a strictly local derivative expansion. This suggests that local mesonic EFTs, while useful for organizing the low-momentum structure, do not by themselves provide the most robust representation of the charged quasiparticle pole in the magnetic-field range relevant for the lattice turnover. 

More generally, the results show that in strong magnetic fields the notion of hadron “mass” becomes intrinsically scheme dependent, while the physically relevant charged excitation is controlled by pole structure, Landau-level kinematics, and residue behavior taken together. The non-monotonic behavior of the charged pion is therefore not simply a mass shift, but a manifestation of how magnetic fields reshape the quasiparticle content of hadronic modes.

The same lesson is expected to apply beyond the $\pi-\rho$ system. Whenever a magnetic background allows mixing between charged hadrons with the same quantum numbers, the resulting spectrum may depend sensitively on how the charged pole is defined and normalized. The framework developed here therefore provides a useful starting point for studying analogous effects in other channels, such as kaons and their vector partners.

\section{Outlook}
\label{sec_outlook}
The mechanism identified here, mixing between charged hadronic modes with the same quantum numbers in a magnetic background, amplified by the suppression of quasiparticle residues, is expected to be more general than the $\pi^\pm-\rho^\pm_{s_z=0}$ system studied in this work. A natural extension is the $K^\pm-K^{*\pm}$ system, where lattice simulations also indicate non-monotonic behavior. It would be particularly interesting to examine whether the same hierarchy found here persists there as well, namely whether the turnover remains robust only in Landau projected pole-based extractions while becoming weaker in strictly local descriptions.

On the lattice side, the present analysis points to several concrete directions. Since the turnover is most naturally associated with the charged Landau projected quasiparticle pole, it would be valuable to study not only the lowest charged pseudoscalar mode but also the corresponding vector channel in the same magnetic field range. In particular, a more direct information on residue, or on the behavior of vector channel correlators in the longitudinal sector, would provide an important test of the residue enhanced mixing mechanism suggested here.

Several improvements of the present framework are also worth pursuing. On the microscopic side, it would be important to compute the full quark-level $\rho\pi\gamma$ triangle vertex in the magnetic field, so that the magnetic dependence of the mixing can be determined more directly rather than split into an explicit loop part plus a local matched remainder. It would also be valuable to include finite-width effects of the rho meson in a more systematic way, in order to test how stable the pole-based picture remains once the vector mode acquires a sizable imaginary part.

More broadly, the present study suggests that strong magnetic fields reshape hadron spectroscopy not only through shifts of mass parameters, but through changes in the very definition of the relevant quasiparticle excitation. Understanding this interplay between Landau-level kinematics, pole structure, and residue suppression may therefore be essential for connecting lattice QCD, effective field theory, and microscopic quark models in magnetized QCD matter.

\acknowledgments
The author would like to thank Pengfei Zhuang, Shijun Mao, Hengtong Ding for fruitful discussion. The work is supported by NSFC grant Nos. 12005112 and The Fundamental Research Funds for Beijing Municipal Universities.

\appendix
\section{Matsubara sum in the loop}
\label{App_matsubara}
The functions $I_2(q_0)$ and $I_3(q_0)$ encodes the Matsubara sum, 
\begin{eqnarray}
\label{I2I3q0}
I_2(q_0)&=&i\int \frac{dk_0}{2\pi}\frac{-1}{((k_0+q_0)^2-E_{u,n}^2)(k_0^2-E_{d,m}^2)}\nonumber\\
&=&\frac{1}{4E_{u,n}E_{d,m}}\Big(\frac{n_F(E_{d,m})-n_F(E_{u,n})}{(E_{d,m}-E_{u,n}+q_0)}
+\frac{n_F(E_{d,m})-n_F(E_{u,n})}{(E_{d,m}-E_{u,n}-q_0)}\nonumber\\
&&\qquad\qquad 
+\frac{1-n_F(E_{d,m})-n_F(E_{u,n})}{(E_{d,m}+E_{u,n}-q_0)}
+\frac{1-n_F(E_{d,m})-n_F(E_{u,n})}{(E_{d,m}+E_{u,n}+q_0)}\Big),\nonumber\\
I_3(q_0)&=&i\int\frac{dk_0}{2\pi}\frac{k_0(k_0+q_0)}{((k_0+q_0)^2-E_{u,n}^2)(k_0^2-E_{d,m}^2)}\nonumber\\
&=&\frac{1}{4}\Big(\frac{n_F(E_{d,m})-n_F(E_{u,n})}{(-E_{d,m}+E_{u,n}-q_0)}
+\frac{n_F(E_{d,m})-n_F(E_{u,n})}{(-E_{d,m}+E_{u,n}+q_0)}\nonumber\\
&&+\frac{1-n_F(E_{d,m})-n_F(E_{u,n})}{(E_{d,m}+E_{u,n}-q_0)}
+\frac{1-n_F(E_{d,m})-n_F(E_{u,n})}{(E_{d,m}+E_{u,n}+q_0)}\Big),
\end{eqnarray}
with $E_{u,n}^2=2n|q_uB|+k_3^2+m_f^2$ and $E_{d,m}^2=2m|q_dB|+k_3^2+m_f^2$. 

The Matsubara sum in the loop in parameters 
\begin{eqnarray}
I_{0d}^l&=&\int\frac{dp_E}{2\pi}\frac{1}{p_E^2+E_d^2}=\frac{1-2 n_F\left(E_d\right)}{2E_d},\nonumber\\
I_{0u}^l&=&\int\frac{dp_E}{2\pi}\frac{1}{p_E^2+E_u^2}=\frac{1-2 n_F\left(E_u\right)}{2E_u},\nonumber\\
I_{1}^l&=&\int\frac{dp_E}{2\pi}\frac{1}{(p_E^2+E_u^2)(p_E^2+E_d^2)}=\frac{1-2 n_F\left(E_u\right)}{2 E_u \left(E_d^2-E_u^2\right)}-\frac{1-2 n_F\left(E_d\right)}{2 E_d \left(E_d^2-E_u^2\right)},\nonumber\\
I_{2}^l&=&\int\frac{dp_E}{2\pi}\frac{1}{(p_E^2+E_u^2)^2(p_E^2+E_d^2)}=\frac{1-2 n_F\left(E_d\right)}{2 E_d \left(E_d^2-E_u^2\right){}^2}+\frac{\left(E_d^2-3 E_u^2\right) \left(1-2 n_F\left(E_u\right)\right)}{4 E_u^3
   \left(E_d^2-E_u^2\right){}^2}+\frac{n'_F\left(E_u\right)}{2 E_d^2 E_u^2-2 E_u^4},\nonumber\\
I_{3}^l&=&\int\frac{dp_E}{2\pi}\frac{1}{(p_E^2+E_u^2)^3(p_E^2+E_d^2)}=\frac{n''_F\left(E_u\right)}{8 E_u^3 \left(E_u^2-E_d^2\right)}+\frac{\left(3 E_d^2-7 E_u^2\right) n'_F\left(E_u\right)}{8 E_u^4 \left(E_d^2-E_u^2\right){}^2}\nonumber\\
&&\qquad\qquad\qquad\qquad\qquad\qquad\quad -\frac{1-2
   n_F\left(E_d\right)}{2 E_d \left(E_d^2-E_u^2\right){}^3}-\frac{\left(-10 E_d^2 E_u^2+3 E_d^4+15 E_u^4\right) \left(1-2 n_F\left(E_u\right)\right)}{16 E_u^5
   \left(E_u^2-E_d^2\right){}^3}.
\end{eqnarray}

\section{Landau level parameters $H_i$ in the loop integral }
\label{App_H}
The parameters in \eqref{RPAkernal} and \eqref{bosonaction} are obtained through loop functions which all contain integral over perpendicular momentum. Such integral are translated into the following integrals over Laguerre polynomials
\begin{eqnarray}
H_1^{n,m}&=&\int du~e^{-\frac{3u}{2} }\Big(L_m(2u)L_{n-1}(u)+L_n(u)L_{m-1}(2u)\Big),\nonumber\\
H_2^{n,m}&=&\int du~e^{-\frac{3u}{2} }u~L^1_{m-1}(2u)L^1_{n-1}(u),\nonumber\\
H_3^{n,m}&=&\int du~e^{-\frac{3u}{2} }\Big(L_m(2u)L_{n-1}(u)-L_n(u)L_{m-1}(2u)\Big),\nonumber\\
H_4^{n,m}&=&\int du~e^{-\frac{3u}{2} }\Big[L_m(2u)\Big((u-2)L_{n-1}(u)+4(u-1)L_{n-2}^1(u)+4u L_{n-3}^2(u)\Big)\nonumber\\
&&+L_{m-1}(2u)\Big((u-2)L_{n}(u)+4(u-1)L_{n-1}^1(u)+4u L_{n-2}^2(u)\Big)\Big]\equiv\mathcal{F}_{n-1,m}+\mathcal{F}_{n,m-1},\nonumber\\
H_5^{n,m}&=&\int du~e^{-\frac{3u}{2} } u L_{m-1}^1(2u)\Big((u-4)L_{n-1}^1(u)+4(u-2)L_{n-2}^2(u)+4u L_{n-3}^3(u)\Big).
\end{eqnarray}
These integrals can all be carried out analytically using the following basic integral 
\begin{eqnarray}
\label{basic_integral_1}
J_{n,m}^a&=&\int du~e^{-\frac{3u}{2} }u^aL^a_{m}(2u)L^a_{n}(u)=\frac{\Gamma(m+n+a+1)}{m!n!}\frac{(-1)^m2^{1+a}}{3^{m+n+a+1}}F[-m,-n,-m-n-a,9],
\end{eqnarray}
as well as the relations between Laguerre functions
\begin{eqnarray}
u L_l^a(u)&=&-(a+l) L_{l-1}^a(u)+(a+2 l+1) L_l^a(u)-(l+1) L_{l+1}^a(u),\nonumber\\
L_m^{a-1}(x)&=&L_m^a(x)-L_{m-1}^a(x).
\end{eqnarray}
The analytical expression for $H_i$ defined above are then 
\begin{eqnarray}
H_1^{n,m}&=&J_{m,n-1}^0+J_{m-1,n}^0,\nonumber\\
H_2^{n,m}&=&J_{m-1,n-1}^1,\nonumber\\
H_3^{n,m}&=&J_{m,n-1}^0-J_{m-1,n}^0,\nonumber\\
\mathcal{F}_{n-1,m}
&=&(2n-3)J^{0}_{n-1,m}-nJ^{0}_{n,m}-(n-1)J^{0}_{n-2,m}\nonumber\\
&&+4(2n-3)(J^{1}_{n-2,m}-J^{1}_{n-2,m-1})
-4(n-1)(J^{1}_{n-1,m}-J^{1}_{n-1,m-1})
-4(n-1)(J^{1}_{n-3,m}-J^{1}_{n-3,m-1})\nonumber\\
&&+4(2n-3)( J^{2}_{n-3,m}-2J^{2}_{n-3,m-1}+J^{2}_{n-3,m-2})\nonumber\\
&&-4(n-2)(J^{2}_{n-2,m}-2J^{2}_{n-2,m-1}+J^{2}_{n-2,m-2})-4(n-1)(J^{2}_{n-4,m}-2(n-1)J^{2}_{n-4,m-1}+J^{2}_{n-4,m-2}),\nonumber\\
H_5^{n,m}&=&\Big[(2n)J^{1}_{n-1,m-1}-nJ^{1}_{n,m-1}-nJ^{1}_{n-2,m-1}-4J^{1}_{n-1,m-1}\Big]\nonumber\\
&&+4\sum_{r=0}^{n-2}\Big[(2r+2)J^{1}_{r,m-1}-(r+1)J^{1}_{r+1,m-1}-(r+1)J^{1}_{r-1,m-1}-2J^{1}_{r,m-1}\Big]\nonumber\\
&&+4\sum_{r=0}^{n-3}(n-2-r)\Big[(2r+2)J^{1}_{r,m-1}-(r+1)J^{1}_{r+1,m-1}-(r+1)J^{1}_{r-1,m-1}\Big].
\end{eqnarray}

\section{Landau-projected polarization function}
\label{App_project}
Take the Landau-projected pion polarization function $\Pi_{\pi\pi,j}$ in \eqref{chargedloop} as an example, we briefly show the procedure of the calculation. 
\begin{eqnarray}
\label{Pipipi}
\Pi_{\pi\pi,j}&=&-i2N_c\int_{r,r'}\text{Tr}\Big[\varphi^*_{j}(r')(i\gamma^5)S_u(r,r')(i\gamma^5)S_d(r',r)\varphi_{j}(r)\Big]\nonumber\\
&=&-2iN_c\int d^2 r_\perp d^2 r'_ \perp\int\frac{d^4q}{(2\pi)^4}\frac{d^4k}{(2\pi)^4} [2\pi\delta(p_0+q_0-k_0)]^2[2\pi\delta(p_3+q_3-k_3)]^2\times \text{tr}[S_{u}(q)i\gamma^5S_{d}(k)i\gamma^5]\nonumber\\
&&\times e^{i \vec{q}_\perp\cdot(\vec{r}_\perp-\vec{r}'_\perp)+i \vec{k}_\perp\cdot(\vec{r}'_\perp-\vec{r}_\perp)}e^{-i\frac{2}{3}\frac{(x-x')(y+y')}{2l^2}}e^{i\frac{1}{3}\frac{(x'-x)(y+y')}{2l^2}}\varphi^*_{jp_1}(\vec{r}\,'_\perp)\varphi_{jp_1}(\vec{r}_\perp).
\end{eqnarray}
$S_{u}$ and $S_{d}$ are the Fourier transformation of the translation invariant part of the quark propagator. $\varphi_{np_1}$ are the wave function of the Landau level defined in \eqref{wavefunction}. Consider the lowest Landau level $n=0$ in the external line, inserting the explicit expression of the Landau level wave function, and combining with the Schwinger phase in the quark propagator, the integral over $\vec{r}_\perp$ and $\vec{r}'_\perp$ can be performed, giving 
\begin{eqnarray}
\label{integralyy}
\int dxdx' \int dydy' e^{i \vec{q}_\perp\cdot(\vec{r}_\perp-\vec{r}'_\perp)+i \vec{k}_\perp\cdot(\vec{r}'_\perp-\vec{r}_\perp)}e^{-i\frac{(x-x')(y+y')}{2l^2}}\varphi^*_{jp_1}(\vec{r}\,'_\perp)\varphi_{jp_1}(\vec{r}_\perp)
&=&2\pi 2l^2e^{-l^2(\vec{q}_\perp-\vec{k}_\perp)^2}.
\end{eqnarray}
Inserting this back to \eqref{chargedloop}, it becomes 
\begin{eqnarray}
\label{quarkloop1}
\Pi_{\pi\pi,j=0}&=&-2iN_c{2\pi} 2l^2\int\frac{d^4q}{(2\pi)^4}\frac{d^4k}{(2\pi)^4} [2\pi\delta(p_0-q_0+k_0)]^2[2\pi\delta(p_3-q_3+k_3)]^2e^{-l^2(\vec{q}_\perp-\vec{k}_\perp)^2} \text{tr}[\widetilde{S}_{u}(q)i\gamma^5\widetilde{S}_{d}(k)i\gamma^5].
\end{eqnarray}
The Dirac trace in \eqref{Pipipi} is evaluated as 
\begin{eqnarray}
\text{Tr}[{S}_{u}(q)i\gamma^5{S}_{d}(k)i\gamma^5]
&=&e^{-q_\perp^2l_u^2-k_\perp^2l_d^2}\sum_{n,m=0}^\infty\frac{(-1)^{n+m}}{[q_0^2-2n|q_uB|-q_3^2-m_f^2][k_0^2-2m|q_dB|-k_3^2-m_f^2]}\nonumber\\
&&\times\big\{ 8(k_0q_0-k_3q_3-m_um_d)[L_n(2q_\perp^2l_u^2)L_{m-1}(2k_\perp^2l_d^2)+L_{n-1}(2q_\perp^2l_u^2)L_{m}(2k_\perp^2l_d^2)]\nonumber\\
&&+64\vec{k}_\perp\cdot\vec{q}_\perp L_{n-1}^1(2q_\perp^2l_u^2)L_{m-1}^1(2k_\perp^2l_d^2)\big\}
\end{eqnarray}
Pluging in the Dirac trace and $\vec{r}_\perp\vec{r}'_\perp$ integral into \eqref{Pipipi}, and carrying out the integral over $q_0$ and $q_3$, one arrives at 
\begin{eqnarray}
\label{quarkloop2}
\Pi_{\pi\pi,j=0}&=&-2iN_c{2\pi}  2l^2\int\frac{d^2q_\perp d^4k}{(2\pi)^6}e^{-l^2(\vec{q}_\perp-\vec{k}_\perp)^2}e^{-\frac{3}{2}q_\perp^2l^2}e^{-3k_\perp^2l^2}\nonumber\\
&&\times\sum_{n,m=0}^\infty\frac{(-1)^{n+m}}{G_{u,n}G_{d,m}}
\big\{ 8(k_0q_0-k_3q_3-m_um_d)[L_n(3q_\perp^2l^2)L_{m-1}(6k_\perp^2l^2)+L_{n-1}(3q_\perp^2l^2)L_{m}(6k_\perp^2l^2)]\nonumber\\
&&\qquad\qquad\qquad\qquad+64\vec{k}_\perp\cdot\vec{q}_\perp L_{n-1}^1(3q_\perp^2l^2)L_{m-1}^1(6k_\perp^2l^2)\big\},
\end{eqnarray}
where $G_{u,n}=(p_0+k_0)^2-2n|q_uB|-(p_3+k_3)^2-m_f^2$ and $G_{d,m}=k_0^2-2m|q_dB|-k_3^2-m_f^2$. The integral over $\vec{q}_\perp$ in \eqref{quarkloop2} can be carried out analytically using 
where we have used the integral 
\begin{eqnarray}
\int_0^\infty x^{\nu+1} e^{-\beta x^2} L_n^\nu(\alpha x^2) J_\nu (x y) dx=2^{-\nu-1}\beta^{-\nu-n-1}(\beta-\alpha)^n y^\nu e^{-\frac{y^2}{4\beta}} L_n^\nu[\frac{\alpha y^2}{4\beta(\alpha-\beta)}].
\end{eqnarray}
Inserting back to \eqref{quarkloop2}, one has ($p\rightarrow q$, and work in the $q_z=0$ limit)
\begin{eqnarray}
\label{quarkloop3}
\Pi_{\pi\pi,j=0}&=&-4iN_c\int\frac{d^4k}{(2\pi)^4}\sum_{n,m=0}^\infty\frac{(-1)^{m}}{G_{u,n}G_{d,m}}\frac{8}{  5^{n}}\Big\{(k_0(p_0+k_0)-k_3^2-m^2)\Big[e^{-\frac{18}{5}k_\perp^2l^2}\Big(\frac{1}{5}L_n(-\frac{12}{5}k_\perp^2l^2) L_{m-1}(6k_\perp^2l^2)\nonumber\\
&&\quad-L_{n-1}(-\frac{12}{5}k_\perp^2l^2)L_{m}(6k_\perp^2l^2)\Big)\Big]
-\frac{16}{5}k_\perp^2e^{-\frac{18}{5}k_\perp^2l^2}L_{n-1}^1(-\frac{12}{5}k_\perp^2l^2)L_{m-1}^1(6k_\perp^2l^2)\Big\},
\end{eqnarray}
the integral over $\vec{k}_\perp$ can also be carried out analytically using \eqref{basic_integral_1}, yielding $Y_1$ and $Y_2$. One can then arrive at the polarization $\Pi_{\pi\pi,0}$ in \eqref{polarization}. For other Landau-projected polarization function, there is another coefficient $Y_3$. We present the definition of $Y_i$ below
\begin{eqnarray}
Y_1(m,n)&\equiv&\frac{(-1)^{m}}{5^{n+1}}\int du e^{-\frac{3}{5}u}\Big[L_n(-\frac{2}{5}u)L_{m-1}(u)
 -5L_{n-1}(-\frac{2}{5}u)L_{m}(u)\Big]
 =-\frac{2^{m-1}}{3^{n+m}}\frac{(2n+m)\Gamma(n+m)}{\Gamma(1+n)\Gamma(1+m)},\nonumber\\
Y_2(m,n)&\equiv&\frac{(-1)^{m}}{5^{n+1}}\int du e^{-\frac{3}{5}u} u L_{n-1}^1(-\frac{2}{5}u)L_{m-1}^1(u)
=-\frac{ 2^{m-1}}{3^{n+m}}\frac{\Gamma(n+m)}{\Gamma(n)\Gamma(m)},\nonumber\\
Y_3(n,m)&\equiv&\frac{(-1)^{m}}{5^{n+1}}\int du e^{-\frac{3}{5}u}\Big[L_n(-\frac{2}{5}u)L_{m-1}(u)+5L_{n-1}(-\frac{2}{5}u)L_{m}(u)\Big] =-\frac{2^{m-1}}{3^{n+m}}\frac{(m-2n)\Gamma(n+m)}{\Gamma(1+n)\Gamma(1+m)}.
\end{eqnarray}

\bibliographystyle{unsrt} 
\bibliography{ref}

\end{document}